\documentclass[range]{ar2e}
\begin{document}

\input epsf.tex    

\input psfig.sty

\jname{Ann. Rev. Astron. Astrophys.}
\jyear{2003}
\jvol{50}
\ARinfo{1056-8700/97/0610-00}

\title{Enhanced Angular Momentum Transport in Accretion Disks}

\renewcommand{\baselinestretch}{1.00}
\markboth{Balbus}{Angular Momentum Transport in Disks}

\author{Steven A. Balbus
\affiliation{Department of Astronomy, VITA,
University of Virginia, 
Charlottesville, VA 22901\ sb@virginia.edu}}


\begin{abstract}
The status of our current understanding of angular momentum transport
in accretion disks is reviewed.  The last decade has seen a dramatic
increase both in the recognition of key physical processes and in our
ability to carry through direct numerical simulations of turbulent
flow.  Magnetic fields have at once powerful and subtle influences
on the behavior of (sufficiently) ionized gas, rendering them directly
unstable to free energy gradients.  Outwardly decreasing angular velocity
profiles are unstable.  The breakdown of Keplerian rotation into MHD
turbulence may be studied in some numerical detail, and key transport
coefficients may be evaluated.  Chandra observations of the Galactic Center
support the existence of low luminosity accretion, which may ultimately
prove amenable to global three-dimensional numerical simulation.

\end{abstract}

\maketitle

\small\normalsize
\parskip = 0.5em

\newcommand\bb[1] {   \mbox{\boldmath{$#1$}}  }
\newcommand\vecc[1] {   \mbox{\boldmath{$#1$}}  }
\newcommand\vvec[1] {   \mbox{\boldmath{$#1$}}  }

\newcommand\apj{ {\it Ap. J.\ }}
\newcommand\ApJ{ {\it Ap. J.\ }}
\newcommand\mn{ {\it MNRAS\ }}
\newcommand\asap{ {\it Astron.\ Astrophys.\ }}

\newcommand{\dd}{\partial}
\newcommand{\del}{\mbox{\boldmath{$\nabla$}}}
\newcommand{\bcdot}{\mbox{\boldmath{$\cdot$}}}
\newcommand{\btimes}{\mbox{\boldmath{$\times$}}}
\newcommand{\bcalFE}{\mbox{\boldmath{${\cal F_E}$}}}
\newcommand{\bcalFJ}{\mbox{\boldmath{${\cal F_J}$}}}
\newcommand{\kva}{\bb{k\cdot v_A}}
\newcommand{\vv}{\bb{v}}
\newcommand{\ve}{\bb{v_e}}
\newcommand{\uu}{\bb{ u}}
\newcommand{\ww}{\mathbf{ w}}
\newcommand{\iw}{{i\omega}}
\newcommand{\B}{\bb{B}}
\newcommand{\WRphi}{{ \langle u_R u_\phi-u_{A\,R}u_{A\,\phi}\rangle}}
\newcommand{\beq}{\begin{equation} }
\newcommand{\eeq}{\end{equation}}
\newcommand{\wbar}{ \bar{\omega} }
\newcommand{\delH}{ {\cal\delta H} }
\newcommand{\Rel} { {\rm Re} }
\newcommand{\Img} { {\rm Im} }

\newcommand{\schwz}{ {\dd  \ln P\rho ^{-5/3} \over \dd Z}}
\newcommand{\schwR} { {\dd  \ln P\rho ^{-5/3} \over \dd R} }

\newcommand{\balbz}{ {\dd  \ln T \over \dd Z}}
\newcommand{\balbR} { {\dd  \ln T \over \dd R} }

\newcommand{\BV} {Brunt-V\"ais\"al\"a\ }

\vskip 1.2 cm

\parbox{5.5cm}{\it I hate being ``allowed for,'' as if I were 
some incalculable quantity in an astronomical equation.}

\smallskip
\hskip 1 cm {\it -- D.L. Sayers, {\rm The Documents in the Case.}}

\bigskip

\section{Introduction}

In recent years, accretion disk transport theory has developed so
rapidly, any review is destined to be significantly dated the
moment it appears in print.  This will put the reader at a
disadvantage.  However, it is an exhilarating time for disk theorists.

The current happy state of affairs in this computationally-driven field
is largely due to the swift evolution of three-dimensional
magnetohydrodynamical (MHD) codes and their supporting hardware.  These
powerful tools arrive with provident timing, coinciding with a
deepening theoretical understanding of the role of magnetic fields in
accretion disk dynamics.  The result is that accretion disk turbulence
theory has grown from a mere viscosity coefficient to a fully
quantitative science.  In this review, I will focus on what
is now known of the relationship between turbulence and enhanced
angular momentum in accretion disks, and the resulting implications for
some selected astrophysical systems.

The classical problem with accretion disks is that they do of course
accrete.   How is it that fluid elements orbiting in a central force
field lose their specific angular momentum and spiral inwards?  One may
quickly rule out ordinary particulate viscosity.  Astrophysical disks
are simply too big.  To fix ideas, note that disturbances are propagated
by viscous diffusion over a distance $l$ on a time scale of order
$l^2/\nu$, where $\nu$ is the kinematic viscosity, or about $3\times
10^7$ years for $l\sim 10^{10}$ cm and $\nu = 10^5$ cm$^2$ s$^{-1}$.
This is orders of magnitude too long for the time variability seen in
compact object accretion disks.

The way around this difficulty was perceived to be turning the woefully
inadequate viscosity to one's advantage by appealing to the associated
large Reynolds number.   This, it was thought, would turn differential
rotation into shear-driven turbulence (e.g., Crawford \& Kraft 1956,
Shakura \& Sunyaev 1973).  The breakdown of shear flow lamina into
turbulence was known since the work of Reynolds to be
triggered by nonlinear flow instabilities.  The fact that there were no
demonstrable local {\em linear} instabilities for a Keplerian rotation
profile was therefore not immediately viewed as an embarrassment to
this scenario.

Shear turbulence is a desirable trait in a flows where greatly
enhanced (angular) momentum transport is needed.  This is because
shear-driven turbulence is characterized by a high degree of
correlation between the radial and azimuthal velocity fluctuations.  This,
as we shall see, has the direct effect of raising a disk's angular
momentum flux orders of magnitude above what would be possible with
an ordinary collisional viscosity.  

No topic in fluid dynamics is more contentious than the onset and
development of turbulence, and accretion disk turbulence has not been
an exception.   Keplerian disks do not resemble laboratory shear flows,
however locally one peers.  Coriolis forces dramatically stabilize
rotational flow on large scales and small, a feature not shared
by classical {\em planar} Couette flow or Poiseuille flow.\footnote
{Another key feature that disks do {\em not} share with these laboratory
flows is the presence of a boundary layer at a retaining wall.}  Yet
Coriolis forces do no work on the fluid; they are in fact absent from the
energy conservation equations.  Because a potent free energy source in
the form of shear retains a presence with or without Coriolis forces,
debate has centered upon whether fluid nonlinearities at a high enough
Reynolds number would still find a way to tap into this source when
linear disturbances fail to do so.

No published laboratory experiment has shown the breakdown of a
Keplerian-like Couette flow profile.  But truly definitive studies
have yet to be performed.   The onset of nonlinear instabilities that
have been reported in Couette flow experiments are generally related
to very sharp rotational velocity gradients typical of Kelvin-Helmholtz
instabilities, not the $r^{-1/2}$ power law characteristic of a Keplerian
profile (Triton 1988).  The issue is receiving renewed attention, with
groups in Saclay, Los Alamos, and Princeton looking at Reynolds numbers
$ Re \sim 10^{5-6}$, much in excess of those available in the classical
experiments (e.g., Coles 1965).

Theoretical developments have been more brisk.  A theory for the
onset of turbulence in Poiseuille flow was elucidated in the 1970's
and 80's (Bayly, Orszag, \& Herbert 1988).  The key is the existence
of neutrally stable (or slowly decaying) finite amplitude disturbances.
These solutions are time-steady but spatially periodic in the
streamwise direction (Zahn et al.\ 1974).  It is these amended flow
profiles that find themselves subject to a rapid, short-wavelength {\em
three-dimensional} instability, leading to a breakdown into turbulence
(Orszag \& Patera 1980, 1981).  A similar process appears to be at
work in shear layers (Pierrehumbert \& Widnall 1982; Corcos \& Lin
1984).   It is now generally accepted that the triggering of a rapid,
linear three-dimensional instability of a nearly neutrally stable,
two-dimensional, finite-amplitude disturbance is a very generic mode of
the breakdown of laminar flow into turbulence.

How does this bear on our understanding of accretion disks?  The most
important point is that even incompressible axisymmetric disturbances
in a rotating fluid will propagate in the form of linear inertial waves
(Lighthill 1978) with a characteristic frequency proportional to the
local vorticity of the rotation profile.  Thus, in astrophysical disks,
the finite amplitude, neutrally stable axisymmetric state that is critical
to the transition to turbulence in mixing layers cannot form.  Only in
a low-vorticity rotational profile, in which the oscillation frequency
is much smaller than the shearing rate, would we expect a breakdown of
flow lamina into turbulence, similar to what is seen in a mixing layer.
This is in good accord with three-dimensional numerical simulations
(Balbus, Hawley, \& Stone 1996; Hawley, Balbus, \& Winters 1999),
which find no nonlinear local instabilities in Keplerian disks, but do
indeed find that shear layers and low-vorticity disks show a nonlinear
breakdown to turbulent flow.  There is as yet no analytic proof of local
nonaxisymmetric stability, however, and the notion that there may be
another nonlinear hydrodynamical route to turbulence in Keplerian disks,
beyond the highest resolution available in supercomputer simulations,
retains some advocates (Richard \& Zahn 1999).

Accretion disks have one critically important attribute not shared with
the classical hydrodynamical fluids: they are generally magnetized.
By engendering new degrees of freedom in their host fluid, even very weak
magnetic fields completely alter the stability behavior of astrophysical
gases, both rotationally and thermally (Balbus 2001).  Free energy sources
in the form of angular velocity and temperature gradients become directly
available to destabilize the flow.

The counter-intuitive point here is that a weak magnetic field can have
such a potent influence.  The stability behavior of strongly magnetized
disks is rich and astrophysically interesting (e.g.\ Terquem \&
Papaloizou 1996; Varni\`ere \& Tagger 2002), but the emphasis in
this review is decidely on subthermal magnetic fields.  Rather than
depending directly on the strength of the equilibrium magnetic fields,
weak field instabilities depend directly upon hydrodynamic properties
of the unperturbed disk.  If the angular velocity decreases outward
in a weakly magnetized accretion disk, which is generally the case,
the rotation profile is linearly unstable (Balbus \& Hawley 1991).
This instability is known as the {\em magnetorotational instability}, or
MRI for short.  As we shall see, the physics of the MRI is very simple.
Nevertheless, a full understanding of its mathematical generality and
wide applicability was much belated, following phenomenological disk
theory (Shakura \& Sunyaev 1973) by nearly two decades.  Knowledge of the
instability itself significantly predated modern accretion disk theory
(Velikhov 1959), albeit in a rather formal global guise.

The numerical study of the MRI does not require unattainable grid
resolutions, and it can be readily simulated.  Both local (Hawley,
Gammie, \& Balbus 1995; Brandenburg et al. 1995) and global (Armitage
1998, Hawley 2000) investigations unambiguously show a breakdown of
laminar Keplerian flow into well-developed turbulence.  The MRI is the
only instability shown to be capable of producing and sustaining the
enhanced stress needed for accretion to proceed on viable timescales in
non-self-gravitating disks.  At low temperatures and high densities,
e.g., in the outer regions of cataclysmic variables (Gammie \& Menou
1998), or in protostellar disks on AU scales (Gammie 1996) the level
of MRI-induced turbulence can change rapidly, erupting or even turning
off completely.  All this occurs while the underlying Keplerian profile
remains essentially fixed.  In short, the instability seems capable of
the full range of accretion complexity manifested in nature.  For all
these reasons, despite the difficulties of coping with MHD turbulence,
the MRI is now at the center of numerical accretion disk studies.

Let us, however, postpone our discussion of matters magnetic, and turn
our attention first to the study of simple hydrodynamical waves in disks.
These are of great practical interest in their own right, especially in
protoplanetary disks.   But it is also the case that understanding the
transport properties of waves deepens one's understanding of turbulent
transport, both hydrodynamic and magnetohydrodynamic, and that is our
primary reason for reviewing them here. We shall then follow with a
discussion of hydrodynamical instability, with a focus on how global
instability can in principle emerge in a differentially rotating disk
even when the Rayleigh condition is satisfied.  Magnetic instability and
magnetic turbulence are the topics of the next section.  Magnetic stresses
are the most important transport mechanism in non self-gravitating
disks, provided the gas is minimally ionized to couple to the field.
The final two sections are a presentation of recent numerical studies
of MHD turbulence, and a summary follows.

\section {Preliminaries}

\subsection {Fundamental Equations}

For ease of future reference, we list here the fundamental equations
of magnetohydrodynamics.
\begin{equation}\label{mass}
{\dd\rho\over \dd t} + \del\bcdot (\rho {\vv}) =  0
\end{equation}
\begin{eqnarray}\label{mom}
\rho {\dd{\vv} \over \dd t} + (\rho \vv\bcdot\del)\vv = -\del\left(
P + {B^2\over 8 \pi} \right)-\rho \del \Phi + 
\left( {\B\over
4\pi}\bcdot \del\right)\B
\nonumber\\
\hfill
+\eta_V\left( \nabla^2\vv
+{1\over3}\del\left(\del\bcdot\vv\right)\right)
\end{eqnarray}
\beq \label{ind}
{\dd{\B}\over \dd t} = \del\btimes\left( {\vv} \btimes {\B} -
\eta_B \del\btimes{\B}\right)
\eeq
\beq\label{ent}
{P\over\gamma - 1} {d\ln P\rho^{-\gamma}\over dt} = Q^{+} - Q^{-}
\eeq
Equation (\ref{mass}) is mass conservation, (\ref{mom}) is
the dynamical equation of motion,
(\ref{ind}) is the induction
equation, and (\ref{ent}) is the entropy equation.
Our notation is standard: $\rho$ is the mass density, $\vv$
the fluid velocity, $P$ the pressure (plus radiation pressure when
important), $\Phi$ the gravitational potential, $\B$ the magnetic field
vector, $\gamma$ is the adiabatic index, $Q^{+}$ ($Q^{-}$)
represent heat gains (losses),
$\eta_V$ the microscopic kinematic shear viscosity, and
$\eta_B$ the microscopic resistivity. 
The azimuthal component of the equation of motion deserves
separate mention,
as it is a direct expression of angular momentum conservation:
\beq\label{angmom}
{\dd (\rho R v_\phi)\over \dd t} + \del\bcdot\left[
\rho R v_\phi \bb{v} - {R B_\phi\over 4\pi} \bb{B}+
\left(P + {B^2\over 8\pi}\right) \bb{e_\phi}\right] = 0
\eeq
where $\bb{e_\phi}$ is a unit vector in the $\phi$ direction.
The dissipative terms have been dropped because they appear only in the
flux term, transporting a negligible amount of angular momentum.

It is also useful to have at hand an equation for total energy conversation.
This is somewhat lengthy to derive (Balbus \& Hawley 1998), but the result
is readily interpreted:
\beq\label{toten}
{\dd{\cal E}\over \dd t} + \del\bcdot\bcalFE = - Q^-
\eeq
where the energy density ${\cal E}$ is
\beq\label{enden}
{\cal E} = {1\over 2} \rho v^2 + {P\over \gamma - 1} + \rho \Phi +
{B^2\over 8 \pi}
\eeq
and the energy flux is 
\beq\label{enflux1}
\bcalFE= \bb{v} \left( {1\over 2} \rho v^2 + {\gamma P \over
\gamma - 1} +\rho \Phi\right) + {\bb{B} \over 4\pi }
\times(\bb{v\times B}).
\eeq
The energy density consists of kinetic, thermal, gravitational, and
magnetic components; the flux is similar with the magnetic component
present as a Poynting flux.  The heating term $Q^+$, an entropy source,
is assumed to arise from microscopic dissipation, and it does not
explicitly appear in the total energy equation --- it simply converts
one form of energy to another.  The radiative $Q^-$ term, on the other
hand, represents genuine systemic energy losses, and appears
explicitly in the conservation equation. 

\subsection {Nonlinear Fluctuations}

Both waves and turbulence involve the concept of well-defined departures
of the flow from a smooth background.  Velocity fluctuations are of
particular interest, because it is possible to formulate an exact energy
conservation law for the fluctuations themselves.  This, in turn,
explicitly shows the role of differential rotation as a source of free
energy for the (correlated) turbulent fluctuations associated with
outward transport of angular momentum.

Let us define the velocity fluctuation $\bb{u}$ by
\beq
\bb{u} = \bb{v} - R\Omega(R)\, \bb{e_\phi}.
\eeq
$\Omega$ is in principle arbitrary, but of course the
motivation for this definition is that $\Omega(R)$ is a reasonably good
approximation to an underlying rotation profile for the accretion
flow.   It is possible to combine the equation of motion (\ref{mom})
with the internal entropy equation (\ref{ent}) to obtain an exact,
$\phi$ averaged 
energy equation for the $u$ velocity fluctuations alone:
\beq\label{fluc}
{\dd {\cal E}_u\over \dd t} + \del\bcdot\bcalFE_{\bb{u}} = -
\left(\rho\, u_R u_\phi - {B_R B_\phi\over 4\pi} \right)
{d\Omega\over dR} - Q^- .
\eeq
Here ${\cal E}_u$ is the fluctuation energy density
\beq
{\cal E}_u = 
{1\over2}\rho (u^2 + \Phi_{eff}) + {P\over\gamma-1}+ {B^2\over 8\pi},
\eeq
$\Phi_{eff}$ is an effective potential function
\beq
\Phi_{eff} = \Phi - \int^R R\Omega^2\, dR,
\eeq
and $\bcalFE_{\bb{u}}$ is the energy flux of the fluctuations
themselves:
\beq
\bcalFE_{\bb{u}} = \bb{u} \left( {1\over2} \rho u^2 +{\gamma P \over\gamma-1}
+\rho\Phi_{eff} \right) + {\bb{B}\over4\pi} \btimes (\bb{u}\btimes B).
\eeq
Because equation (\ref{fluc}) has been averaged over $\phi$, only $R$ and $Z$
components appear in the flux.  

The combination
\beq
T_{R\phi} = \rho\, u_R u_\phi - {B_R B_\phi\over 4\pi} 
\eeq
is an important quantity in both turbulent and wave transport theories of
accretion disks.  It has appeared once before: within the angular momentum
conservation equation (\ref{angmom}), where it emerges as a component
in the flux term.  Its constituents may be separately identified as
Reynolds ($\rho u_Ru_\phi$) and Maxwell ($- B_RB_\phi/4\pi$) stresses.
(Note that both wavelike and turbulent disturbances can create tight
radial--azimuthal correlations in the velocity and magnetic fields.)
These correlations evidently serve two conceptually quite different
functions: they directly transport angular momentum, and as shown
in equation (\ref{fluc}), they tap into the free energy source of
differential rotation.  The latter role is particularly crucial for
sustaining turbulence.  Without external driving, the only energy source
for the fluctuations is this coupling of the stress to the differential
rotation.  In astrophysical accretion disks that make use of this free
energy source, $T_{R\phi}$ must have the same sign as $-d\Omega/dR$,
i.e, it must be positive.

\section{Hydrodynamic Waves in Disks}

\subsection {The Linear Wave Equation}

Consider an unmagnetized disk in which the pressure and density obey a simple
polytropic equation of state, $P = K\rho^\gamma$, where $K$ is a
constant.  We may define an enthalpy function ${\cal H}$:
\beq
{\cal H} = \int {dP\over \rho} = {\gamma P/\rho\over \gamma - 1} = 
{a^2\over \gamma - 1},
\eeq
where $a^2$ is the adiabatic sound speed.  It is convenient to work
in standard cylindrical coordinates $(R, \phi , Z)$.  The gas rotates in the
gravitational field of a central mass.  
The angular velocity $\Omega$ must be constant on cylinders,
$\Omega = \Omega (R)$ (Tassoul 1978).  

Although it is a standard approximation, the assumption of a barotropic
equation of state is obviously an idealization.  Among other shortcomings,
it precludes the possibility of a buoyant response in the from of
internal gravity waves due to \BV oscillations (Ogilvie \& Lubow 1999).
But in standard disk models, entropy stratification arises because of
radiative heat diffusion from turbulent heating.  A linearized wave
treatment of such an ``equilibrium'' is at best a delicate matter.
In general, the vertical temperature structure of accretion disks is
not well-understood, and the virtue of adopting a barotropic pressure
is that it allows important dynamical behavior to be revealed.

Our goal is to study how linearized wave disturbances
transport energy and angular momentum through a Keplerian
disk.  To this end, we introduce small perturbations to the
equilibrium solution, denoted as $\delta \rho, \, \delta\bb{v}$,
etc.  The equilibrium solution is axisymmetric, so a perturbed
flow quantity $X$ has the form
\beq
\delta X = \delta X (R, Z) \exp (im\phi - i \omega t),
\eeq
where $m$ is an integer and $\omega$ is the wave frequency.
For the moment, the $R, Z$ dependence of the amplitude is unrestricted.
The linearized dynamical equations of motion are
\beq
- i \bar{\omega} \, \delta v_R - 2 \Omega \, \delta v_\phi = 
- {\dd \,\delta{\cal H}\over \dd R}
\eeq
\beq
 - i \bar{\omega} \, \delta v_\phi + {\kappa^2  \over  2\Omega} \, \delta v_R =
 - i {m\over R}\,  \delta{\cal H}
\eeq
\beq
-i \bar{\omega}  \, \delta v_Z = - {\dd \, \delta{\cal H}\over \dd Z}.
\eeq
We have introduced the Doppler-shifted wave frequency,
\beq
\bar{\omega} = \omega - m \Omega,
\eeq
and what is known as the epicyclic frequency $\kappa$:
\beq
\kappa^2 = 4\Omega^2 + {d\Omega^2\over d\ln R}.
\eeq
The epicyclic frequency is the rate at which a point mass in a circular
motion, disturbed in the plane of its orbit, would oscillate about its
average radial location (Binney \& Tremaine 1987).  A negative value
of $\kappa^2$ quite generally implies that axisymmetric disturbances
are hydrodynamically unstable.  The
requirement $\kappa^2 > 0$ is 
known as the Rayleigh stability criterion.

The remaining equations are the linearized mass conservation equation
\beq\label{linmass}
-i \bar{\omega}  \, {\delta\rho\over\rho} + {1\over\rho}
\del\bcdot(\rho\, \bb{\delta v}) = 0.
\eeq
and the equation of state (relating density and enthalpy perturbations):
\beq
\delta{\cal H} = a^2 {\delta\rho\over\rho}
\eeq

The three dynamical equations may be solved for $\bb{\delta v}$ in terms of
$\delta{\cal H}$:
\beq\label{deluR}
\delta v_R = {i\over D} \left[ \bar{\omega} {\dd{\delta \cal H} \over\dd R}
- {2\Omega m\over R} \, \delta {\cal H}\right],
\eeq
\beq\label{deluphi}
\delta v_\phi = {1\over D} \left[
{\kappa^2\over 2\Omega} {\dd\delH\over\dd R} - {m\wbar\over R}\, \delH
\right],
\eeq
\beq\label{deluz}
\delta v_Z = - {i\over \wbar} {\dd\delH\over \dd Z},
\eeq
where 
\beq
D = \kappa^2 -  \wbar^2.
\eeq
Using equations (\ref{deluR} -- \ref{deluz}) in (\ref{linmass})
and simplifying the results, one obtains the linear wave equation for the disk:
\beq\label{waveeq}
\left[
{1\over R}{\dd\ \over\dd R}\left( {R\rho\over D}{\dd\ \over\dd
R}\right) - {1\over \wbar^2}{\dd\ \over\dd Z}\left(
\rho{\dd\ \over \dd Z} \right) - 
{m^2\rho \over R^2 D} + {1\over R\wbar} {\dd\ \over\dd R} 
\left( 2 \Omega m \rho\over D \right) +{\rho\over \epsilon^2a^2} \right]\delH = 0
\eeq
We have inserted an artificial $\epsilon$ factor in the sound speed
term, which, though formally equal to unity, will be used as an
aid for sorting out asymptotic orders in a WKB analysis.
(The disk is here presumed to be cold in the sense that $R\Omega \gg a$.)  

The locations at which $D=0$ and $\wbar = 0$ are singularities of
the wave equation, though in the case of the former the singularity
is only apparent, not real.  They are known respectively as Lindblad
and corotation resonances, and their neighborhoods are zones where
waves couple strongly to the disk.  They are of importance in the
study of tidally driven waves, and are critical to an understanding
of planetary migration (Goldreich \& Tremaine 1979, 1980; Ward 1997).
In this section, our emphasis will be upon freely propagating WKB waves,
and we shall assume that neither $D$ nor $\wbar$ is small; {\em i.e.,}
that we are not in the neighborhood of resonance.

\subsection{Two Dimensional WKB Waves}

\subsubsection {First Order: Dispersion Relation and Group Velocity}

We seek solutions of equation (\ref{waveeq}) having the form
\beq\label{wkb}
\delH = A(R, Z) \, \exp\left[i  S(R, Z)\over \epsilon \right]
\eeq
The idea is that the phase $S/\epsilon$ varies rapidly, and the $\epsilon$
factor ensures this in the formal limit $\epsilon \rightarrow
0$.   We will solve the wave equation to leading and second order in a
$1/\epsilon$ expansion.  Note that the absolute phase is not relevant
here, and we may assume that $A$, the amplitude, is real. 
We shall also assume that the waves are tightly wound, i.e. that
both $k_R$ and $k_Z$ $ \gg m/R$.

Inserting (\ref{wkb}) into (\ref{waveeq}), we find that the leading order
$1/\epsilon^2$ terms give
\beq\label{disp}
{k_Z^2\over\wbar^2} + {k_R^2\over \wbar^2 - \kappa^2} = {1\over a^2},
\eeq
where
\beq
(k_R, k_Z) = \left( {\dd S/\dd R}, {\dd S/\dd Z}\right).
\eeq
This is the dispersion relation for WKB disk waves, and the gradients of $S$ are,
in essence, the wavenumber components.  

Figure (1) is a plot of the constant frequency curves in the $k_Z a$,
$k_R a$ wavenumber plane.  For $\wbar >\kappa$, the iso-$\wbar$ curves
are ellipses; for $\wbar <\kappa$, they are hyperbolae.  These two
different conic sections define the two distinct wave branches, with
very different transport properties.  Indeed, this diagram makes evident
several remarkable features of disk waves.

The elliptical iso-$\wbar$ surfaces correspond to density waves, which
are rotationally modified sound waves (e.g Goldreich \& Tremaine 1979),
and the hyperbolae correspond to inertial waves (Vishniac \& Diamond 1989),
described below.  It is not difficult to show that the ellipses and
hyperbolae always intersect at right angles, so that the curves
are like a conformal mapping.  The practical relevance of
this is that since the wave group velocity $\bb{U}$ is the wavenumber
gradient of $\wbar$,
\beq
\bb{U} = \left( {\dd\wbar\over \dd k_R}, {\dd\wbar\over \dd k_Z}\right),
\eeq
the group velocity direction of the density waves
lies along the inertial wave iso-$\wbar$ curves, and the group
velocity direction of the inertial waves lies along the density
wave curves.  Low frequency disturbances in disks have very different
transport properties from high frequency disturbances, a point to which
we shall return many times.

\begin{figure}[ht]
\centerline{\epsfbox{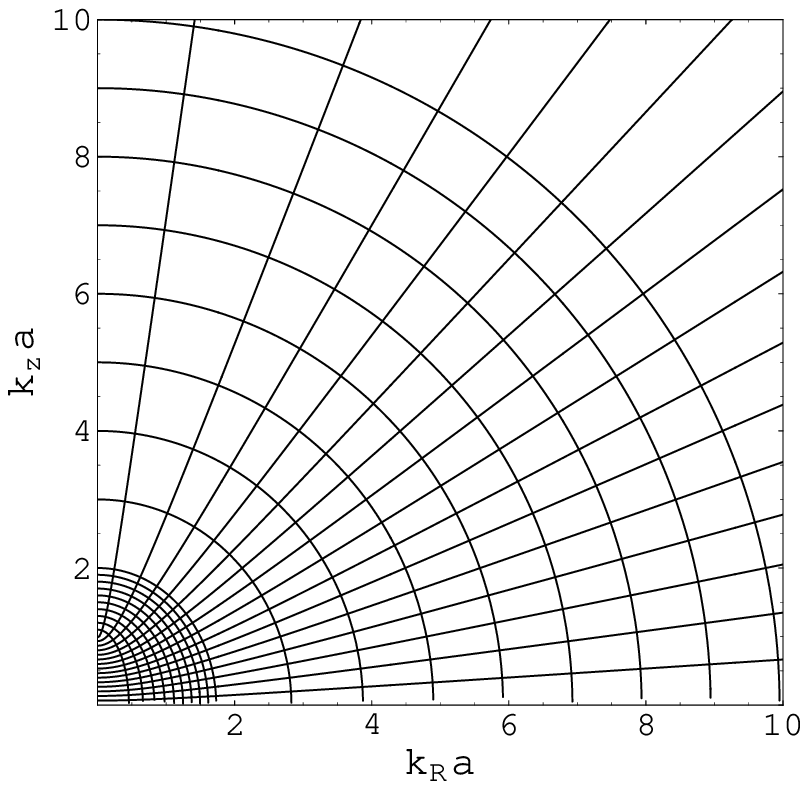}}
\caption{Contours of constant $\wbar$ for the dispersion relation
(\ref{disp}) of a Keplerian disk.  These form a set of conformal ellipses
(density waves) and hyperbolae (inertial waves).  The value of $\wbar$
along a curve is read off at the point of intersection with the $k_R =
0$ axis; the numerical scale is in units of $\Omega = 1$.  Density wave
ellipses are separated by one unit for $\wbar>2$ and 0.1 units for $1.1<
\wbar< 2$.  Inertial wave hyperbolae are separated by .066 units.}
\label{fig1}
\end{figure}

For a given wavevector $(k_R, k_Z)$, the gradient to an
iso-$\wbar$ curve could point in either direction, because
the dispersion relation (\ref{disp}) does not uniquely
determine the sign of $\wbar$.  Direct calculation reveals:
\beq
U_R \equiv {\dd \wbar\over \dd  k_R} = - { k_R/\wbar D \over (k_R/D)^2
+(k_Z/\wbar^2)^2},
\eeq
\beq
U_Z\equiv {\dd \wbar\over \dd k_Z} = { k_Z/\wbar^3\over (k_R/D)^2
+(k_Z/\wbar^2)^2},  
\eeq
and
\beq
\bb{k\cdot U} = \left[ a^2\wbar\left( k_R^2/D^2 +
k_Z^2/\wbar^4\right) \right]^{-1}
\eeq
The sign of the group velocity must be carefully determined; it depends
upon $\bb{k}$, $\wbar$, and $D$.  By way of illustration, assume
that the wavenumber components are all positive.  If $\Omega$ decreases
with increasing $R$, density waves ($D< 0$) beyond the corotation
radius will propagate radially outward, and radially inward inside of
corotation.  They will propagate upwards outside corotation, and
downwards inside corotation.  Inertial waves ($D>0$) propagate in the
{\em opposite} radial sense as density waves, but in the same vertical
direction.  Finally, the epicyclic response $D=0$ is degenerate, and
$k_R$ must vanish at this location, the Lindblad resonance.  The WKB
treatment breaks down here, and other techniques must be used
(Goldreich \& Tremaine 1979).


\subsubsection {Density Waves}

The general dispersion relation can be written in the form
\beq\label{dispalt}
\wbar^4 - (k^2 a^2 +\kappa^2)\wbar^2 + \kappa^2 k_Z^2 a^2 = 0 
\eeq
where $k^2 = k_R^2 + k_Z^2$.  
In this form the density and inertial wave branches separate cleanly, with 
the density wave corresponding to a dominant balance between the first
two terms in (\ref{dispalt}), and inertial waves between the final two terms.
The density wave branch of the dispersion relation (\ref{disp}) is thus,
\beq
\wbar^2 = k^2 a^2 +\kappa^2. 
\eeq
The condition that the final term in the dispersion relation be small
compared with the first two can be met either if $ka\gg\kappa$, in which
case the problem reduces to ordinary acoustic waves, or if $k_Z\ll k_R$,
which is the classical one-dimensional density wave problem.  In either
case, the density wave group velocity $\bb{U}$ tends to be parallel
to $\bb{k}$, and thus perpendicular to constant phase surfaces.  

\subsubsection {Inertial Waves}

Inertial waves may be less familiar than their density wave counterparts
because of the latter's
well-known role in the theory of galactic spiral structure.  Indeed,
as we have just seen, inertial waves disappear entirely in a ``thin''
disk with $k_Z=0$, which doubtless contributed to their undeserved
second-class role.  It must be emphasized that inertial waves are the
{\rm only} avenue of response available to a polytropic disk at driving
frequencies less than $\kappa$, and the properties of these waves are
worthy of attention.

The inertial wave branch cleanly separates from the density wave branch
in the limit
$ k_Z a \gg \kappa $, and is essentially incompressible. 
Inertial waves satisfy the dispersion relation (\ref{dispalt})
with the dominant balance attained in the last terms,
obtained in the $a\rightarrow \infty$ limit: 
\beq\label{inert}
\wbar^2 = {k_Z^2\over k_R^2 + k_Z^2} \kappa^2.
\eeq
(Note that WKB inertial waves do not require $R\Omega  \gg a$.)
Equation
(\ref{inert}) tells us that only the vertical component of the wave
number contributes to the inertial response.  This is readily
understood: since incompressible fluid displacements are nearly
perpendicular to their wavevectors, it is only the planar
component of the {\em displacement} that elicits an inertial response.  It
is the inertial Coriolis force that provides the return
impetus in this class of wave; such forces are accessed only by
displacements in the disk plane.

For an inertial wave, $\bb{k\cdot U}$ will typically be of order
$\kappa^3/k_Z^2 a^2$, and is thus very small, $ \bb{k\cdot U}  \ll kU$.
Since $\bb{k}$ is by definition perpendicular to constant phase surfaces,
the inertial wave group velocity lies entirely within these same surfaces.
Unlike density waves, which radiate ringlike from their point of origin,
inertial waves have the unusual property that they exude, tube-like,
slowly establishing their pattern.  Driven incompressible turbulence
in an otherwise stable hydrodynamic disk is basically an ensemble of
interacting inertial waves.

\subsubsection {Second Order: Wave Action Conservation}

We move to the next order in our WKB analysis.  If we gather the terms of
order $1/\epsilon$ that arise from substituting (\ref{wkb}) into
(\ref{waveeq}), we find that they may be combined into the form of an exact
divergence:
\beq\label{action}
{\dd\ \over \dd Z} \left[{\rho k_Z A^2\over
\wbar^2}\right]  
- {1\over R} {\dd\ \over\dd R}\left[R\left(\rho k_R A^2\over D\right)\right] 
=0,
\eeq
or
\beq
\del\bcdot\left[ \rho A^2 \wbar\left( k_R^2/D^2 +
k_Z^2/\wbar^4\right) \bb{U}\right] = 0.
\eeq
This expresses the conservation of wave action, and the quantity
multiplying the group velocity vector $\bb{U}$ is thus (up to an overall
sign) the wave action density.  In the next section, we will show that
the wave action is proportional not to the wave energy, which is not
a conserved quantity (unless $m=0$), but to the wave angular momentum.
Except at resonances, this is always conserved in a dissipationless fluid.

\subsection{Angular Momentum and Energy Wave Fluxes}

The wave angular momentum flux is given by
\beq
\bcalFJ = \rho  R \left \langle \Rel\left(\delta v_\phi\right)\,
\Rel\left(\bb{\delta v}\right) \right\rangle
\eeq
where the angle brackets denote an average over one wave cycle,
and $\Rel$ denotes the real part.  (Hereafter, in all expressions
quadratic in the $\delta$ wave amplitudes, it is understood that the
real part is to be taken, unless explicitly written otherwise.)
Note that if the
waves could impart a mean mass flux $\dot m$ to the background, such
terms would produce a contribution to the angular momentum flux of the
form ${\dot m}Rv_\phi$.  However, the dissipationless
passage of a density or inertial wave does not 
cause matter to flow in the disk, despite the fact that
$$
\langle\delta\rho\, \delta v_R \rangle \not= 0.
$$
This is because the perturbed quantities have been calculated only to linear
order in their amplitudes, whereas the fluxes of interest
are {\em second order} in the same amplitudes.  Thus, the perturbed velocity
is not really $\bb{\delta v}$, but $\bb{\delta v} + \bb{v_2}$, where
$\bb{v_2}$ includes
all higher order corrections 
in the velocity amplitude.  These need not be wave-like.  It is
only when $\bb{v_2}$ is included, making a second order contribution
to the mass flux as a product with the equilibrium density $\rho_0$,
that the mass flux disappears:
$$
\langle\delta\rho\, \delta v_R\rangle  +
\langle \rho_0 v_{2R}\rangle = 0 .
$$
Fortunately, $\bb{v_2}$ itself does not have to be computed explicitly.
Instead, one simply requires $\langle\rho\bb{v} \rangle$ to vanish, while
retaining all other leading order terms.   These will be quadratic in
the linear amplitudes; no other second order perturbations subsequently
appear.  (This is {\em not} true of dissipational flow, turbulent or
wavelike, where mass accretion is clearly a possibility.)  In the end,
we shall see that all wave fluxes of interest are directly proportional
to a conserved wave action.


To determine more precisely the relationship between
the wave action and angular momentum,
let us calculate the radial component of
the angular momentum flux.  
Using equations (\ref{deluR}) and (\ref{deluphi}),
it is straightforward to show that the radial component of the angle
momentum flux is 
\beq
\rho R \langle \delta v_R\, \delta v_\phi \rangle
= - {m \rho k_R\over D} \,
{A^2\over 2} = {m\rho k_R\over \wbar^2 - \kappa^2} \, {A^2\over 2}.
\eeq
The factor of $1/2$ comes from averaging over a wave cycle.  Reference to
equation (\ref{action}) shows that
this is 
exactly the radial component of the wave action, multiplied by the
proportionality constant $m/2$.  It is often
useful to express this flux in terms of radial velocity fluctuations.
This can be done with the help of
equation (\ref{deluR}), neglecting the $m$ term in the WKB limit:
\beq\label{Rang}
\rho R \langle \delta v_R\, \delta v_\phi \rangle
= \rho {m\over k_R}\, \left( 1 - {\kappa^2\over \wbar^2} \right) \langle (\delta
v_R)^2\rangle. 
\eeq
The $Z$ component of the angular momentum flux need not be
directly computed, but instead can be obtained by multiplying the
radial flux component by the wave group velocity ratio $U_Z/U_R$:
\beq\label{Zang}
\rho R \langle \delta v_Z\, \delta v_\phi \rangle
={m\rho k_Z\over \wbar^2} {A^2\over 2} = {m\rho\over k_Z} \langle (\delta
v_Z)^2\rangle, 
\eeq
and therefore
\beq\label{actbis}
\rho R\langle \bb{\delta v} \, \delta v_\phi \rangle = {m\rho
A^2\over 2} \left( -{k_R\over D}, {k_Z\over \wbar^2}\right).
\eeq
Equation (\ref{actbis}), together with wave action
conservation (\ref{action}), ensure that angular momentum 
flux is conserved,
\beq\label{waveact}
\del\bcdot\left[ \rho R \langle \bb{\delta v} \, \delta v_\phi
\rangle\right] = 0, 
\eeq
as embodied in (\ref{angmom}).  

How does this bear on energy conservation?  Although the total
mechanical energy of the disk-wave system must be conserved (absent
dissipation), wave energy is {\em not}, in general, conserved.  
The rate at which wave energy is exchanged with the disk's differential 
rotation is in fact 
\beq\label{sbs}
{\rm volumetric\ energy\ exchange\ rate} = - \rho\langle \delta v_R\, 
\delta v_\phi \rangle {d\Omega\over d\ln R}, \eeq
{\em i.e.,} the exchange rate is ``stress by strain,'' as per equation 
(\ref{fluc}).  An explicit calculation of this is revealing, however.

Equation (\ref{enflux1}) and
the requirement of vanishing mass flow leave
\beq\label{enflux}
\bcalFE  = 
\left( {1\over 2} \rho v^2 +{\gamma P \over \gamma - 1 } \right)
\bb{v}
\eeq
as the contributing energy flux terms. 
The first term is the contribution from the background, and it vanishes
in a comoving frame.  The second, representing the wave itself, is
the enthalpy flux.  Once again, since no contribution to the enthalpy
flux
can come directly from mass flow, only one term quadratic in
the wave amplitudes survives the averaging procedure:
\beq\label{avgenflux}
\langle P\bb{v} \rangle = 
\langle \rho \left({P\over \rho} \right) \bb{v} \rangle \rightarrow
\rho \left\langle \delta \left({P\over\rho}\right) \, \bb{\delta v}
\right\rangle.
\eeq
Now, for a polytropic equation of state,
\beq
\delta\left(P\over\rho\right) = {\gamma -1\over\gamma}{\delta P\over
\rho},
\eeq
hence the wave (enthalpy) flux is just
\beq
\bcalFE\ {\rm (wave)} = 
\langle \delta P\, \bb{\delta v} \rangle = 
\rho \langle \delta {\cal H} \, \bb{\delta v} \rangle.
\eeq
This is a general and well-known result that can be derived in
other ways (e.g.\ Lighthill 1978).
From equations (\ref{deluR})--(\ref{deluz}),
we use the leading order WKB terms, which combine to give the $(R, Z)$ components
of the flux:
\beq
\rho \langle \delta {\cal H} \, \bb{\delta v} \rangle = 
{\wbar \rho A^2\over2} \left( -{ k_R\over D}, {k_Z\over \wbar^2}
\right) = 
{\wbar\over m} \rho R \langle \delta v_\phi \, \bb{ \delta v}
\rangle
\eeq
This is the mechanical analogue of the wave
quantum condition: the wave energy is the product of the wave frequency
and the quantized angular momentum unit.  Note that the relevant
frequency is Doppler shifted relative to the moving medium,
as a comoving observer
in the disk would measure it.  
Wave action conservation 
(\ref{waveact}) now gives directly
\beq
\del\bcdot \left[\rho \langle \delta {\cal H} \, \bb {\delta v} \rangle
\right]=
 {d(\wbar/m)\over d R} \rho  R  \langle \delta v_\phi \, \delta v_R
 \rangle =
- \rho \langle \delta v_\phi \, \delta v_R
\rangle {d\Omega\over d\ln R},
\eeq
which shows that the wave energy grows at a 
rate given by the stress-by-strain formula (\ref{sbs}).

To verify that the total energy of the disk and wave system is conserved,
we return to the energy flux (\ref{enflux}).  Using
``no mass flux'' reasoning similar to
that used for determining the enthalpy wave flux, the kinetic
energy flux becomes
\beq
{1\over2} \rho v^2 \bb{v}  \rightarrow \rho R \Omega \langle
\delta v_\phi \, \bb{ \delta v} \rangle,
\eeq
so that it, too, is just proportional to the angular momentum flux. 
The net energy flux of the wave plus the background kinetic energy
flux is
\beq
\rho R \langle \delta v_\phi \, \bb{\delta v} \rangle 
\left({\wbar\over m} +
\Omega \right) =
{\omega\over m} \rho R \langle \delta v_\phi \, \bb{ \delta v} \rangle 
\eeq
Since $\omega$ is constant, wave action (or angular momentum) conservation 
guarantees that this quantity, the total energy flux, is 
likewise conserved.   

The following summary statements can be made:
(1) Wave action and wave angular momentum flux are
both conserved quantities.  (2) 
The wave energy flux is related to wave angular momentum flux by
\beq
\bcalFE\ {\rm (wave)} = {\wbar\over m} \bcalFJ = 
{\wbar\over m} \rho R \langle \delta v_\phi \, \bb{ \delta v} \rangle,
\eeq
and is {\em not}  conserved.  (3) The total energy flux 
of the wave plus background is related to the
wave angular momentum flux by 
\beq
\bcalFE\ {\rm (total)} = {\omega\over m} \bcalFJ =
{\omega\over m} \rho R \langle \delta v_\phi \, \bb{ \delta v} \rangle,
\eeq
and {\em is} conserved.   Wave action, angular momentum flux, and
total energy flux are all essentially the same thing, differing
from one another only by constant proportionality factors.

\subsection{ Wave Angular Momentum Transport in Protoplanetary Disks.}

In any shearing disk, local nonaxisymmetric disturbances tend to evolve
toward a trailing configuration: a wave crest advances in $R$ as it
decreases, or trails, in $\phi$.  This is an obvious consequence of an
outwardly decreasing rotation profile; leading configurations are
created when the angular velocity increases outward.  Since $m d\phi +
k_R dR = 0$ along a constant phase path, $m$ and $k_R$ must have the
same sign in a trailing wave form.  From this result and equation
(\ref{Rang}), it follows that {\em trailing density waves transport
angular momentum radially outward, while trailing inertial waves
transport angular momentum radially inward.}  This is a conclusion of
some astrophysical significance.

One of the most interesting and important applications of wave
angular momentum transport occurs in protoplanetary disks.   
An planet embedded in such a disk will tidally excite density waves
in regions where $\wbar^2 > \kappa^2$.  (Goldreich \& Tremaine 1979,
1980).  Though the wave group velocity and energy flux will differ
in sign on either side of the planet's orbit, the angular momentum 
flux will always be directed outwards.  There are then two questions
that immediately arise: (1) Will the planet lose enough angular
momentum to the waves that it will eventually be accreted by the
central star?  (2) Is this an effective angular momentum transport
mechanism for the disk?

Whether the planet loses or gains angular momentum depends upon
the symmetry of the angular momentum transfer.  The locations at
which $D$ vanishes, the Lindblad resonances, are the launching points
of the density waves.  The fundamental frequency of the driving
potential is of course just the planetary orbit frequency $\Omega_p$,
but there are Fourier components at all integer multiples $\omega = m
\Omega_p$.  Hence the Lindblad resonances are defined by 
$$\Omega_p -  \Omega = \pm \kappa/m$$ or
\beq
r = \left( m\pm 1 \over m\right)^{2/3}r_p,
\eeq
where $r_p$ is the planet's location.  The outer resonances thus tend to
be somewhat closer to the planet than the inner resonances (Goldreich \&
Tremaine 1980, Ward 1986) with the consequence that angular momentum is
lost to the disk, and the planet migrates inwards.  The time scale for
orbital evolution is embarrassingly rapid: if the present day Jupiter
were at its current location in a disk at a temperature of 100 K and a
column density of 100 g cm$^{-2}$,
its orbit would evolve in a few thousand years!

Once the interaction between the planet and the mass becomes nonlinear,
material within the planet's Roche lobe is efficiently evacuated,
and a gap forms (Lin \& Papaloizou 1979).  Material inside the planet
loses angular momentum and falls in, whereas material outside the planet
acquires angular momentum and moves out.  The subsequent evolution of the
planet then follows that of the disk itself, and the planet moves in on
the timescale of the accretion process.  In standard disk models (Hartmann
et al. 1998), this also is rapid compared with the disk's lifetime.

It is not an easy matter to go beyond these simple estimates, and the
problem of how to halt migration is a serious one.  Torques exerted at
the location of the planet itself (corotation) could conceivably help, but
these are difficult to evaluate, as are the complications associated with
magnetic couplings, should these be present (Nelson \& Papaloizou 2003).
It is also possible that a gap may open in a quiescent disk region near
the midplane, while accretion occurs via a magnetically active envelope
(Gammie 1996).

The question of whether an ensemble of planets embedded in a disk might
serve as an effective viscosity has been examined recently by Goodman \&
Rafikov (2001).  The idea is that density waves would dissipate locally,
but also transport energy and angular momentum like a kinematic viscosity.
Goodman \& Rafikov argue that if all the dust in the solar nebula
were gathered into an ensemble of Earth-like planets, the effective
Reynolds number (using the sound speed and a vertical scale height) of
the planet-gas fluid would be $10^3$--$10^4$.  As noted by these authors,
however, even Earth-mass planets migrate rapidly, and are also thought to
form only late in the disk's evolution.  The nature of angular momentum
transport in protostellar/planetary disks remains very much open.

\subsection {Angular Momentum Transport and Weak Turbulence}

For a given mass and angular momentum, the lowest energy configuration
of a disk is not a disk at all, but a singular state in which all the
mass is at the center and all the angular momentum is at infinity
(borne by a negligible amount of mass).  The ubiquitous appearance
of trailing density waves in spiral galaxies is understood as
as an attempt to reach this minimum energy state (Lynden-Bell \& Kalnajs
1972).  

Now self-gravitating density waves, unlike those considered here, are
not restricted to values of $\wbar>\kappa$.  But without self-gravity,
{\em any} trailing wave disturbance with a comoving frequency less
than the local epicyclic frequency will drive angular momentum ``the wrong
way,'' inward, and will not move the disk toward its ultimate minimum
energy state.  This statement depends only upon the wave's trailing
nature; the group velocity, for example, could have either sign.
Only trailing, compressive disturbances transport angular momentum
outward, and absent a powerful organizer like self-gravity, these are
expensive to excite and maintain.

Whether one envisions hydrodynamical disk turbulence as an ensemble
of interacting waves, or more phenomenologically as dissipative mixing
of fluid elements, the tendency is for angular momentum to go inwards.
To understand the case for the latter, consider the mixing of two different
fluid elements, $A$ and $B$, with $A$ originating farther out in the disk.
When they meet at some intermediate radius, $A$ will tend to have higher
angular momentum than $B$, since angular momentum increases outward in
a Rayleigh-stable disk.  When it is at the same radius as $B$, $A$ will
necessarily have a higher angular velocity, and angular momentum will be
transported from $A$ to $B$ if the mixing lowers the energy (Balbus 2000).
In other words, dissipative mixing transports angular momentum inwards.
(Note that rigorous specific angular momentum conservation by the
displaced fluid elements is not essential to this argument.)

This wave behavior may be the explanation behind the finding of numerical
simulations showing that local inward angular momentum transport generally
ensues when turbulence in Keplerian disks is driven externally (Stone
\& Balbus 1996, Cabot 1996), but that vigorous disturbances can in
fact drive outward angular momentum transport.  We have just argued
that direct turbulent transport from fluid element mixing is itself
likely to result in inward transport.  It is also likely, however, that
unstable convective modes will mix with and excite stable wave modes,
since only a rather small subset of wavenumbers are formally unstable.
Incompressible convection couples more strongly to inertial waves than
to density waves.  The resulting sheared, trailing wave forms would
transport angular momentum inwards.  By way of contrast, vigorously
driven convection on sufficiently rapid timescales (faster than the
sound crossing time over a vertical scale-height), can couple directly to
compressive {\em density waves,} whose trailing configurations transport
angular momentum outwards.  Note that thermal convection does not itself
carry an ``inward transport'' label, a point that has been misunderstood.
Any agitation in an incompressible Keplerian disk will likely result in
inward turbulent transport.  Moreover, if trailing wave disturbances
bear the brunt of the angular momentum transport in a nonmagnetized
disk, the sign of the radial flux depends only upon the ratio of the
epicylic frequency to the wave frequency.  Long period, incompressible
disturbances transport angular momentum inwards; rapid, compressible
disturbances transport outwards.

\subsection {Bending Waves}

Radial excursions from a circular orbit cause fluid elements to
oscillate about their equilibrium point at the epicyclic frequency.
A similar frequency $\kappa_Z$ exists for vertical excursions, 
\beq
\kappa_Z^2 = {\dd^2\Phi\over \dd Z^2}
\eeq
where $\Phi$ is the gravitational potential at the unperturbed
(midplane) position of the fluid element (Binney \& Tremaine 1987).
In conjunction with pressure and rotational forces, these can appear
as waves at radial wavelengths much in excess of the disk thickness,
and are a possible source of angular momentum transport in an otherwise
inviscid disk (Papaloizou \& Terquem 1995).   The large $m/Rk_R$ ratio
of these ``bending waves'' suggests that they may transport efficiently.
Moreover, $m=1$ modes in non self-gravitating Keplerian-like disks have
been shown by Papaloizou \& Lin (1994) to propagate with a non-dispersive
group velocity.

Papaloizou \& Terquem (1995) calculated the angular momentum transport
associated with these waves in binary systems in which the companion
does not orbit in the disk plane.  Like density waves, bending waves
grow in amplitude as they approach the center of a disk.   This is
just conservation of wave action, but it is more intuitively grasped
if one thinks of angular momentum conservation and a shrinking radius.
If the amplitude becomes nonlinear, it will tend to dissipate, so that
the effect is a sort of nonlocal disk viscosity.  Typical accretion
times in binary systems were found to be at least a few times $10^7$
years, at the upper end of estimated disk lifetimes (Hartman 1998).
If noncoplaner tidal interactions are common, warps in protostellar
disks may influence their evolution.

\section {Hydrodynamical Instability}

In the Introduction, we discussed the history of invoking high
Reynolds number shear turbulence as a source for enhanced accretion
disk viscosity. These invocations ignore the stabilizing influence
of a positive angular momentum gradient.  It is occasionally argued
that this is just a matter of scale, that stabilizing Coriolis forces
are less effective at very small scales, and that the appearance of
nonlinear instability requires higher Reynolds numbers to observe than
are obtainable with current numerical codes (Longaretti 2002).  But there
are no defining scales when the shear is treated as homogeneous and the
Coriolis parameter is treated as a local constant.  The stabilization is
completely scale-free, which is why both inertial wave propagation, and
the Rayleigh stability criterion are independent of wavenumber magnitude
(cf. eq.~[\ref{inert}]).  Any local behavior evinced on small scales in
an inviscid shear flow should be seen at larger scales, whether or not
Coriolis forces are present.

The most detailed study of three-dimensional local Keplerian stability
is that of Hawley, Balbus, \& Winters (1999).  Both ZEUS and PPM codes
(which have very different dispersion properties) were run on the same
initial sets of finite-amplitude disturbances, and complete convergence
was achieved.  Resolutions up to $256^3$ grid points were followed.
No evidence of nonlinear instability was found.

By way of contrast, shear layers and local constant angular momentum
disks, which should exhibit nonlinear instability on all scales,
were reported by Hawley et al.\ (1999) to do just that, even at crude
resolutions.  This study is compelling evidence that Keplerian disks are
not prone to local nonlinear shear instabilities.  The results of the
last section show that all localized internal disturbances in such disks
propagate as WKB waves.  Global perturbations that include boundary
dynamics can behave quite differently, however, leading not to wave
propagation, but to true instability---even if there is no violation of the
Rayleigh stability criterion.

Here we will look at the simplest manifestation of global instability,
normally a rather complicated process.  This class of instability, first
studied by Papaloizou \& Pringle (1984, 1985), and later elucidated
by Goldreich, Goodman, \& Narayan (1986), emerges when trapped waves
on either side of a corotation region are able to communicate with one
another.  Typically, on either side of the divide, nonaxisymmetric waves
will carry opposite signs in their energy densities: one raises the local
disk energy density, the other lowers it.  If energy is lost from the
negative energy wave and picked up by its positive energy counterpart,
{\em both} amplitudes grow, and as the energy transfer rate increases,
the makings of an instability are at hand.

One way to incorporate global physics into a disk problem
is to make the boundaries practically local. We therefore consider a radially
thin disk with no vertical structure, rather like a thin pipe or
drinking straw.  The central radius is $R_0$, the radial variable
is $x = R-R_0$.  We may approximate
a power law angular velocity profile $\Omega \sim
R^{-q}$ as locally linear.  With $\Omega_0 = \Omega(R_0)$, the angular velocity is 
\beq
\Omega = \Omega_0( 1  - qx/R_0)
\eeq
over the radial extent of our structure, $R = R_0 \pm s$. 
We retain only the leading order terms in $s/R$.

The equilibrium solution satisfies
\beq
-R\Omega^2(R)  = -{1\over\rho}{dP\over dR}  - {GM\over R^2}
\eeq
To leading order, the equilibrium equation becomes
\beq\label{entgrad}
{1\over\rho}{dP\over dR}= 
{d{\cal H}\over dx} = - (2q-3) \Omega^2 x  
\eeq
where ${\cal H}$ is the enthalpy function.  

Consider Eulerian perturbations to the equilibrium, which are
denoted by $\delta$.  We use the locally Cartesian azimuthal variable
$dy = R d\phi$, and seek plane wave solutions of the form
\beq
\bb{ \delta u} (x, y, t) = \bb{ \delta u}(x) \exp[ik(y -R\Omega_0 t)
-i\omega t ]
\eeq
where $k$ is the azimuthal wavenumber.  We shall consider incompressible
disturbances only.  The linearized equations of motion
are then mass conservation,
\beq
{d(\delta u_x)\over dx} + i k\, \delta u_y = 0
\eeq
and the equations of motion,
\beq
-i(\omega +q\Omega kx)\,  \delta u_x - 2 \Omega \, \delta u_y = - {d(\delta{\cal H})
\over dx}
\eeq
\beq\label{ikdH}
-i(\omega +q\Omega kx)\, \delta u_y + 
{\kappa^2 \over 2 \Omega} \, \delta u_x = - i k \delta{\cal H}
\eeq
These may be combined to give a simple equation for $\delta u_x$:
\beq {d^2(\delta u_x)\over
dx^2} - k^2 \, \delta u_x = 0.
\eeq
The eigensolutions are thus of the form
\beq \delta u_x = C\cosh(kx) + S \sinh(kx).  \eeq

The boundary condition for our problem is that the Langrangian
pressure perturbation must vanish at each free surface.  Since
$\rho$ is constant, this becomes
\beq
\Delta {\cal H} = \delta {\cal H} + \xi {d{\cal H}\over dx} = 0
\eeq
where $\xi$ is the radial displacement at the boundary,
\beq
\xi = {\delta u_x\over -i(\omega +q\Omega kx)},
\eeq
and $\delta {\cal H}$ is given by the azimuthal equation of motion
(\ref{ikdH}).
The boundary condition may then be written
\beq
(\omega +q\Omega kx)^2 {d(\delta u_x)\over dx} + \left[
k (\omega +q\Omega kx) {\kappa^2 \over 2 \Omega} + k^2 {\cal H}'
\right] \delta u_x = 0,
\eeq
where ${\cal H}'$ is the equilibrium enthalpy gradient,
(\ref{entgrad}).
This constraint,
which must be applied at $x=\pm s$,
embodies the real dynamical content of the problem.
Carrying through this exercise to leading order in $ks$
and solving for $\omega$ leads to the
dispersion relation
\beq\label{ppdisp}
\omega^4 - \Omega^2 \omega^2 + 3 (3-q^2)k^2 s^2 \Omega^4 = 0.
\eeq
In the small $k$ limit, the unstable root of (\ref{ppdisp}) is
\beq
\omega = i ks \sqrt{3(q^2-3)}  \Omega, \quad k> 0.
\eeq
Unstable modes are present if $q >\sqrt{3}\simeq 1.732$, whereas the
local Rayleigh criterion for instability is $q > 2$.  Thus, the interplay
between pressure and rotation present in boundary edge modes makes it
possible for them to tap the free energy of differential rotation as a
source of instability in a regime where local WKB modes cannot.

Note that the important case of Keplerian flow ($q=1.5$) is stable.
This is a consequence of the fact that the pressure gradient vanishes for
this rotationally supported flow.  Furthermore, it can be shown that the
same type of calculation, applied to free shear flow (i.e. no Coriolis
forces) produces {\em no} instabilities.  Though normally a stabilizing
influence, pressure gradients and rotation are destabilizing here.
Tuning the right combination of rotation and pressure at the boundaries
allows modes to extract angular momentum (and therefore energy) from
the inner edge of the cylinder and transport it to the outer edge.
In our simple example, it is a straightforward exercise to show that the
unstable eigenmodes correspond to outward angular momentum transport,
whereas the decaying eigenmodes correspond to inward transport.


We have worked this example through in some detail, because it is
nicely illustrative of what is required for hydrodynamical instability
driven by differential rotation.  The mode is nonlocal, and depends upon
radial wave trapping: in the case presented, the modes have no where to go
but the edges.  More complicated configurations, in which a region with
large entropy gradients is embedded in a more uniform Keplerian
disk, show similar unstable behavior (Lovelace et al.\ 1999, Li et
al.\ 2000).  Even when the local Rayleigh criterion is satisfied ($q< 2$),
differential rotation can be destabilized when $q> \sqrt{3} = 1.732$.
Communication between the different parts of the disturbance on either
side of the corotation point ($x=0$ in the example) is essential; it is
easily shown that instability is not present when there is only one edge.
The extraction of free energy from the differential rotation is a delicate
matter depending on the phase correlation between the radial and azimuthal
velocity fluctuations.

The role of Papaloizou-Pringle instabilities in astrophysical disks
has yet to be established.  It is likely to be of secondary importance in
magnetized disks because of the MRI, but it is is possible that there
may be protostellar applications in regions where the disk gas is poorly
coupled to the magnetic field.  Even here, however, large departures
from simple Keplerian profiles are required for instability, and it is
not obvious that these are sustainable.

Global simulations of constant angular momentum tori (Hawley 1991) show
that the nonlinear resolution of the Papaloizou-Pringle instability is not
turbulence, but a spiral pressure wave.  Moreover, the instability rapidly
saturates when even a small amount of accretion is present (Blaes 1987).
As a practical matter, the instability seems to be most important for
nonmagnetized constant angular momentum tori, whereas Keplerian disks are
more or less immune.

\section {MHD Turbulence}

\subsection {Introduction}

In elucidating the density wave theory of spiral structure, Lynden-Bell \&
Kalnajs (1972) posed a simple and revealing question: why should stellar
disks be unhappy without spiral structure?  The answer is that there
are other configurations that share the same mass and angular momentum,
but with less total energy.  These correspond to a concentration of
mass near the disk center, and a transference of angular momentum to
large radial distances.  The point is that in an asymptotic sense, all
of the angular momentum can reside in none of the mass, and, even more
importantly, bearing none of the {\it energy.}  The hard part is to get
the angular momentum out, which is precisely what a trailing density
wave does.  

The disruption of a differentially rotating flow into an accretion
flow, the signature process of this review, occurs when (1) it is
energetically favorable to do so; and (2) a path to these lower energy
states is available.  The lowering of disk energy must be accompanied
by the outward transference of angular momentum, for this is the only
route self-consistent with this objective.  Recall that turbulence
must be actively maintained, because it is highly dissipative: it
needs access to a ready source of free energy.  It is the end goal of
minimized energy that makes the undertaking of this passage through
turbulent flow worthwhile.  The final outcome is a singular state with
all the mass, shorn of its angular momentum, collected at the center
of the potential well.  An accretion disk is really nothing more than
the sequence of intermediate states the gas is forced to pass through
enroute to this final state.

There is a tradition, still ongoing, of assaulting Keplerian disks with
a variety of instabilities taken from the fluid literature, hoping that
one will take.  Although many disk models are indeed vulnerable to one
particular instability or another, this is generally because they have
been so constructed.  Only one process directly exploits the fact that
an accretion disk is fundamentally unsatisfied in a state of Keplerian
rotation: the magnetorotational instability.  It is extraordinarily
effective because magnetic fields are able to tap directly into the free
energy sources of a fluid, the gradients of the flow that would be gone
in an extremum energy state.  It is of some interest to note that this is
no less true of a {\em thermal} free energy source (temperature gradient)
than of the rotational free energy source (angular velocity gradient)
(Balbus 2001).  The exquisite sensitivity to these free energy gradients
imparted to a fluid by a magnetic field fundamentally alters the behavior
of astrophysical gases in ways that hydrodynamical modeling cannot hope
to capture.  Viscous modelers beware.

\subsection{The Magnetorotational Instability: A Simple Overview}

The destabilizing role of a magnetic field in a differentially rotating
fluid can be grasped with the aid of mechanical analogy.  Imagine two
nearby fluid elements displaced in the orbital plane from another by a
spring-like force.  Although nominally binding the elements together,
the spring, if it is weak, will have exactly the opposite effect in a
differentially rotating system.  The element orbiting at smaller radius
rotates more rapidly, and the spring torque lowers its angular momentum;
precisely the reverse happens for the element orbiting at larger radius,
whose angular momentum is correspondingly increased.  That must
mean, however, that the small radius element drops down to yet smaller radii
to accommodate its reduced angular momentum; conversely the large
radius element moves to yet more distant radii.  The spring tension
grows with increasing element separation, and the process runs away.
Note that angular momentum transport is not some nonlinear consequence
of the instability, it is the very cause of the (linear) instability.
It flows, of course, outward.

It has been emphasized that the spring must be weak.  Weak compared
with what?  The answer can be seen by noting our implicit use of 
circular orbit concepts in the above discussion.  This is justified
only if the natural oscillation frequency of the spring is less than
the orbital frequency of fluid elements.  

The simplest fluid system displaying this spring-like instability is an
axisymmetric gas disk in the presence of a weak vertical magnetic field.
The field has no effect on the disk equilibrium, which is a balance of
gravitational and rotational forces.  If a fluid element is displaced
in the orbital plane by an amount $\bb{\xi}$, with spatial dependence
$e^{ikZ}$, the induction (``field-freezing'') equation gives
\beq
\delta \bb{B} = ikB \bb{\xi}, 
\eeq
and the magnetic tension force is then
\beq
{ikB\over4\pi\rho} \, \delta \bb{B} = - (\kva)^2 \bb{\xi}.
\eeq
This is exactly the form of a spring-like force, linearly proportional
to the displacement. 

To follow the consequences of this force, we go into a frame of reference
comoving with a small patch of the disk rotating at angular
velocity $\Omega(R_0)$, where $R_0$ is our fiducial radius.
We focus on a small radial neighborhood, so that
$x = R - R_0$ is small in magnitude compared with $R_0$.  Terms of 
order $1/R$ are ignored, except $\Omega (R) = v_\phi/R$ which
is our clock.  (In this sense, $v_\phi$ is large.)  
In the corotating frame, Coriolis and centrifugal forces are present,
and the difference between the centrifugal and gravitational
forces is
\beq
R\Omega^2 (R_0) - R\Omega^2(R) = - x {d\Omega^2\over d R}
\eeq
to leading order in $x$.  The simplest case to examine is
pressure-free displacements in the disk plane with a
vertical wavenumber.  With $x$ and $y$ respectively representing local
radial and azimuthal displacements,
the equations of motion then take on a very simple form
\beq \label{springr}
{\ddot x} - 2 \Omega {\dot y } = - \left( {d\Omega^2\over
d\ln R} + (\kva)^2 \right) x
\eeq
\beq\label{springaz}
{\ddot  y } + 2 \Omega {\dot  x } = -(\kva)^2  y.
\eeq
These equations are in fact the rigorous leading order WKB equations
for vertical wavenumber in a magnetized disk.

Equations (\ref{springr}) and (\ref{springaz}) have solutions
for $x$ and $y$ 
with time dependence $\exp(-i\omega t)$ provided that the dispersion relation
\beq\label{dispmri}
\omega ^4 - \omega^2 \left[\kappa^2 + 2 (\kva)^2\right] +
(\kva)^2\bigl[ (\kva)^2 + {d\Omega^2\over d\ln R} \bigr] = 0
\eeq
is satisfied.  This is a simple quadratic equation for $\omega^2$, and it
is straightforward to show that there are no unstable roots if and only if
\beq\label{mri}
{d\Omega^2\over dR} \ge 0 
\eeq
which replaces the Rayleigh stability criterion of outwardly increasing angular momentum
(Balbus \& Hawley 1991).  The fastest growing mode has a growth rate
of
\beq
|\omega_{max}| = {1\over2} \left|d\Omega\over d\ln R \right|
\eeq
which occurs at a wave number given by
\beq
\bb{(\kva)}_{max}^2  =  \left( {1\over 4} + {\kappa^2\over 16\Omega^2}
\right) \left| d\Omega^2\over d\ln R\right|.
\eeq
Flows in violation of (\ref{mri}), which include most astrophysical cases
of differential rotation, are prone to the magnetorotational instability.

Note that the maximum growth rate is, in the parlance of galactic
structure, the local Oort-A value of the disk.  Balbus \& Hawley (1992)
conjectured that this is the maximum growth rate that any local shearing
instability can achieve in a disk, a result recently established for a
rather general class of viscoelastic fluids (Ogilvie \& Proctor 2003).

\subsection{The Magnetorotational Instability: A Brief History}

The understanding that the characteristic accretion disk combination of
Keplerian rotation and a magnetic field is highly unstable emerged very
belatedly, only a little more than a decade ago.  Why this is
so is an interesting question, given that disk stability research was
already some 30 years old at the time, and modern accretion disk theory
had been around for more than 20 years.  The MRI calculation is not a
difficult one, and work on the topic was initiated by no less a personage
than Chandrasekhar (1953).  How could such an important
instability have been overlooked for so long?  In fact, it wasn't.
But neither was it well-understood.  

The early studies by Chandrasekhar were extremely formal.  Indeed, his
1953 paper was not at all astrophysical in context; it was a theoretical
Couette flow analysis.  Chandrasekhar limited his work on hydrodynamic
and hydromagnetic stability to those configurations in which a {\em
static} equilibrium could be defined and analyzed globally.  But the
lack of a coherent physical explanation of the instability hampered its
understanding, making it difficult to perceive both its local character
and widespread generality.

Recently proposed MRI laboratory experiments (Ji, Goodman, \& Kageyama
2001; Noguchi et al. 2002; R\"udiger \& Shalybkov 2002) have led to
a detailed reexamination of Chandrasekhar's analysis of dissipative
Couette flow, and rather surprisingly, there appears to have been a rather
straightforward oversight (Goodman \& Ji 2002).  Magnetized shear flow
will of course draw out an azimuthal field from any radial component
that happens to be present. In Chandrasekhar's discussion, the term
responsible for this behavior was dropped from the induction equation, on
the grounds of a small magnetic Prandtl number approximation.\footnote{The
magnetic Prandtl number is the dimensionless ratio of the viscous to
resistive diffusivities.} This was most unfortunate, because the process
represented by this dropped term need not vanish in this limit; indeed,
it is critical for the MRI to function.

How widespread this misapprehension became is difficult to say.
(The error made its way into Chandrasekhar's classic text [1961].)
Much of the fluid and stellar community was aware of the instability,
however, and its curious behavior of ostensibly changing the Rayleigh
criterion discontinuously.  Noteworthy studies include those of Velikhov
(1959, the first derivation of a dissipationless angular velocity
stability criterion for Couette profiles); Chandrasekhar (1960, a
generalization of Velikhov's result), Newcomb (1962, a variational
approach); Fricke (1969, stellar differential rotation), Acheson \&
Hide (1972, geophysical applications); and Acheson \& Gibbons (1978,
more general stellar applications).  But the robustness and the very
general tendency of subthermal magnetic fields of {\em any} geometry to
destabilize flows was never explicitly discussed.  Indeed, the common
wisdom was that magnetic fields were thought to be a stabilizing
influence, and they were certainly a computational nuisance.  It is
telling that there are no references in the accretion disk literature
to either Velikhov (1959) or Chandrasekhar (1960) prior to 1991.

The key conceptual point, which really has been grasped only in the
last decade, is that {\em the limit $B\rightarrow 0$ retains the most
salient behavioral features of a magnetized fluid,} and can be profitably
investigated.  One should think of an accretion disk not as a rotating
fluid with a magnetic field, but as a magnetized fluid with rotation.
By tethering fluid elements, magnetic fields impart {\em dynamical}
significance to free energy gradients, which otherwise are felt only
through their more ghostly diffusive presence.  The stability properties
of even a barely magnetized fluid are qualitatively different from those
of a nonmagnetized fluid --- a result which holds whether the fluid is
rotating or not.

\subsection {General Axisymmetric Stability Criteria}

Convective and rotational instability are best handled simultaneously,
since the affected mode is a joint buoyant-inertial disturbance.
A rather general calculation is given by Balbus (2001), who considered
the stability of a rotating, stratified plasma.  The angular momentum
and entropy distributions may be arbitrary functions of $R$ and $Z$.
With an eye toward applications involving hot dilute plasmas, a Coulomb
thermal conductivity is assumed to be present.  Heat is permitted
to flow only along magnetic lines of force, a restriction which has
important consequences.  (Radiative conductivity, which would dominate
in stellar interiors, gives rise to very different stability properties.)
The flow is stable to axisymmetric disturbances provided that
\beq\label{hoil1}
-{1\over \rho}(\del P)\bcdot\del\ln T
+ {\dd \Omega^2\over \dd \ln R} \ge 0,
\eeq
\beq\label{hoil2}
\left( - {\dd P\over \dd Z} \right) \, \left(
{\dd \Omega^2\over\dd R} \balbz - {\dd \Omega^2\over\dd Z}\balbR
\right) \ge 0.
\eeq
Note that the second inequality states that the angular velocity should
increase outward along isothermal surfaces for stability.  Without Coulomb
conductivity, the stability conditions are 
\beq\label{hoilb1}
-{3\over 5\rho}(\del P)\bcdot\del\ln P\rho^{-5/3}
+ {\dd \Omega^2\over \dd \ln R} \ge 0,
\eeq
\beq\label{hoilb2}
\left( - {\dd P\over \dd Z} \right) \, \left(
{\dd \Omega^2\over\dd R} \schwz - {\dd \Omega^2\over\dd Z}\schwR
\right) \ge 0,
\eeq
(Papaloizou \& Szuszkiewicz 1992, Balbus 1995),
whereas the classical hydrodynamical H\o iland criteria are
(Tassoul 1978)
\beq\label{hoilad1}
-{3\over 5\rho}(\del P)\bcdot\del\ln P\rho^{-5/3}
+ {1\over R^3} {\dd R^4\Omega^2\over \dd R} \ge 0,
\eeq
\beq\label{hoilad2}
\left( - {\dd P\over \dd Z} \right) \, \left(
{\dd R^4 \Omega^2\over\dd R} \schwz - {\dd R^4\Omega^2\over\dd
Z}\schwR
\right) \ge 0.
\eeq

Comparing these three sets of criteria, we see that the effect of
a magnetic field is to replace gradients of conserved quantities in
axisymmetric flow (angular momentum and entropy) with the free energy
angular velocity and temperature gradients.  The MRI is, in fact,
only one manifestation of a very general principle.  Magnetic torques
ensure that angular velocity gradients are the stability discriminants
for differentially rotating flow; magnetically confined conduction,
when dominant, ensures that temperature gradients are the stability
discriminants for a stratified gas.

\subsection{Low-Ionization Disks}

Our discussion has up to now implicitly assumed that the disk behaves
completely hydrodynamically, or is fully magnetic.  Here we focus on
the transition from one behavior to the other.

A low-ionization gas contains at least three different fluids of interest:
neutrals, ions, and electrons.  Since, as we shall see, an astrophysical
fluid typically becomes magnetically well-coupled at very small ionization
fractions, the inertia resides in the neutrals, which couple collisionally
to the ions more effectively than to the electrons.  The electrons, on
the other hand, bear the brunt of electrical current and so most closely
correspond to the fluid associated with inductive field freezing.  It is
only an approximation to write the induction equation in terms of the bulk
fluid velocity; more properly it is the electron velocity that should
appear.  The electrons are then coupled to the ions by self-induction,
whereas the ions are coupled to the neutrals via collisions.

In a fully-magnetized fluid, all of these couplings
are strong.  More generally, the electron fluid velocity
$\ve$ may be rewritten
\beq\label{ve}
\ve = \vv + (\ve - \bb{v_i})+ (\bb{v_i} - \vv)
\eeq
where $\vv$ is the velocity of the dominant neutrals, and $\bb{v_i}$
the ion velocity.  The electron-ion velocity difference is
just proportional to the current, and leads to Hall electromotive
forces in the induction equation (e.g.\ Wardle \& K\"onigl 1993).
The ion-neutral difference is related to ambipolar diffusion, and
is proportional to the Lorentz force (Shu 1992).  The resulting full
induction equation is a bit daunting, but Balbus \& Terquem (2001) give
useful order-of-magnitude scalings for the relative importance of each
of the terms that appear.  One important and often overlooked conclusion
is that in astrophysical settings, the Hall electromotive force
is at least as important as ohmic dissipation, and can be destabilizing
(Wardle 1999; Balbus \& Terquem 2001).

In protostellar disks, an ionization fraction as small as $10^{-13}$ can
start to induce magnetic coupling on AU scales.  However, even this level may
not be attainable in a dusty disk where grains are the dominant charge
carriers (Umebayashi \& Nakano 1988).  To add to the uncertainty, T Tauri
systems are X-ray sources (Glassgold, Feigelson, \& Montmerle 2000;
Fromang, Terquem, \& Balbus 2002), which introduces a potent time-varying
ionization source.  The role of cosmic rays, another potential ionization
source, in maintaining the MRI in the outer disk layers has been examined
by Gammie (1996).

The magnetic behavior of protostellar disks is extremely uncertain,
largely because neither the disposition of the dust nor the chemical
properties of the gas are well understood.  It seems likely that
there will be evolutionary phases when only the most distant (low density)
and innermost (hot) gas will be magnetized, but it is also
possible that an active X-ray source and agglomerated dust grains may
render much of the disk magnetically well-coupled (Fromang et al.
2002).

Protostellar disks are not the only astrophysical venue where these
issues arise.  The outer regions of CV disks (and possibly other close
binary disks) also appear to be in the low-ionization regime (Gammie \&
Menou 1998).   It has been suggested that Hall electromotive forces may
be of some relevance to understanding the ``low state'' of such systems
(Balbus 2002).  In these objects, dust grain physics is less likely
to be an important complication than it is for protostellar disks.
The magnetic coupling properties of cool AGN disks are discussed by
Menou \& Quataert (2001).

\subsection{Turbulent Transport in Thin Disks}

Differentially rotating disks, in particular Keplerian disks, are
unstable in the presence of a magnetic field to the MRI, which transports
angular momentum outward even in the linear stages of the instability.
Although the disk could alleviate its difficulties by heading toward
a state of uniform rotation, this is not what happens.  Rather, the
excess angular momentum acquired by the outer regions instead allows
them to occupy even more remote orbits where they rotate yet more
slowly.  Correspondingly, the inner disk regions that have lost angular
momentum slide into lower lying orbits, and rotate yet more rapidly.
Far from eliminating outwardly decreasing angular velocity gradients,
the disk instead behaves in such a way as to accentuate such
``unfavorable'' rotation profiles.  The ensuing nonlinear turbulence
is compatible with this linear angular momentum segregation: energy
is lost via turbulent dissipation, and the global minimum energy state
corresponds to a mass concentration at the dynamical center and all the
angular momentum dispersed to infinity.  The dispersed angular momentum
need not, as a matter of principle, have finite mass or finite energy
associated with it.   The disk's struggle to attain this singular energy
minimum is the topic of this section.

\subsubsection {Classical Viscous Disk Theory}

Traditionally, anomalous (read ``turbulent'') disk transport has been
modeled as an enhanced Navier-Stokes viscosity, a technique developed
by Lynden-Bell \& Pringle (1974) that illustrated certain key
energetic and evolutionary processes.  The approach was never meant
to embody the dynamics of accretion disk turbulence, a very important
qualification all too often lost in subsequent applications of the
theory.  To understand how MHD turbulent transport is
related to features of viscous disk theory, begin with
the viscous stress tensor, which is assumed to be dominated by its
$R\phi$ component
\beq
\sigma_{R\phi} =  \nu \rho {d\Omega\over d\ln R},
\eeq 
where $\nu$ is the (turbulent) viscous shear diffusivity.
The height integrated equations of mass and angular momentum
conservation are
\beq\label{pringmass}
{\dd \Sigma\over \dd t} + {1\over R}{\dd R\Sigma {v_R}\over \dd R}
=0,
\eeq
and 
\beq\label{pringmom}
{\dd\over \dd t}(\Sigma R^2\Omega ) + {1\over R} {\dd \over\dd R}
(\Sigma R^3 \Omega v_R -\nu \Sigma R^3 {d\Omega\over dR}) =0,
\eeq
where $\Sigma$ is the integrated disk column density.

Under steady state conditions, the mass flux 
$${\dot M} \equiv - 2\pi R \Sigma v_R $$
and the angular momentum flux must both be constant.
Assuming that the viscous stress vanishes at the inner edge of the disk
$R_0$ leads to
\beq
{\dot M}  \left[ 1 - \left(R_0\over R\right)^{1/2} \right]=3\pi \nu
\Sigma.
\eeq
This completely determined the functional form
of the turbulent viscosity parameter $\nu\Sigma$, which 
can now be eliminated in favor of the more accessible
quantity ${\dot M}$.  In particular, $\nu$
thermalizes the free energy of differential
rotation at a rate per unit area given by
\beq\label{thirteen} Q_e =  {\nu \Sigma \over 2} \left(d\Omega\over d\ln
R\right)^2 = 
{9\over 8} \nu\Sigma \Omega^2.
\eeq
Eliminating $\nu$ then gives
\beq\label{qmdot}
Q_e = {3GM{\dot M}\over 8\pi R^3}\left(  1 - \left(R_0\over
R\right)^{1/2}
\right),
\eeq
the classical $(Q_e , {\dot M})$ relationship (Lynden-Bell \& Pringle
1974).  The additional assumption that the heating is radiated locally by
a thermal surface flux $\sigma T_{eff}^4$ then leads to a surface profile
$T_{eff}(R) \propto R^{-3/4}$, and, it can be shown, a characteristic
$\nu^{1/3}$ spectral flux profile (Lynden-Bell 1969).  At this
point, however, the theorist's luck runs out: except for the
occasional cooperative CV system (Frank, King, \& Raine 2002), real
accretion disks have considerably more complex spectra that are not
well-understood (e.g. Blaes 2002).  Nevertheless, equation (\ref{qmdot})
with $Q_e$ determined observationally, is the cornerstone relation of
disk phenomenology, and is widely used to deduce mass accretion rates.
It has the advantage of being independent of the viscosity parameter,
which suggests a validity that extends beyond viscous disk models.

A time-dependent equation for the evolution of $\Sigma$ emerges by combining
equations (\ref{pringmass}) and (\ref{pringmom}):
\beq
{\dd \Sigma \over \dd t} = {3\over R} {\dd\ \over \dd R} \left[
R^{1/2} {\dd\ \over \dd R}\left(\nu R^{1/2} \Sigma\right) \right]
\eeq
This diffusion-like equation is commonly used in accretion disk modeling,
but requires an {\em a priori} specification of $\nu$ to be implemented.
Nevertheless, it shows explicitly how disks evolve to their singular end
state of mass concentration at the origin and angular momentum dispersal
to infinity.

\subsubsection{MHD Turbulent Transport in Thin, Radiative Disks}

The relationship between the phenomenological disk theory of the previous
section and MHD turbulence has been investigated using a simple statistical
approach (Balbus \& Papaloizou 1999).  The idea is to separate
the Keplerian rotation profile from the net velocity vector, and
focus on the residual velocity $\bb{u}$
\beq 
\bb{v} = R\Omega \bb{{\hat e}_\phi} +\bb{ u}.
\eeq
We assume that the mean accretion drift velocity
is small compared with fluctuation
amplitudes.  The hierarchy is
\beq
|\langle \bb {u} \rangle |^2 \ll \langle u^2 \rangle \ll R^2\Omega^2.
\eeq
A key assumption is that since the disk is thin, we may average
fluctuating quantities over a radial distance $\Delta R$, large compared with a
vertical scale-height $H$, but small compared with $R$,
and thereby formulate a ``local'' disk theory.  (This is sensible only
if $H/R \ll 1$, otherwise a time or ensemble average must be used.)
The theory is thus expressed in terms of evolutionary PDEs
in $R$ and $t$, and the dynamical fluid quantities have been vertically
integrated and azimuthally averaged, and weighted by density:
\beq
\label{average}
\langle X\rangle = 
{1\over 2\pi\Sigma \Delta
R}\int^{\infty}_{-\infty}\int^{R+\Delta R/2} _{R-\Delta R/2}
\int^{2\pi}_0 \rho X \> d\phi\, dR\,  dz, \quad H\ll \Delta R \ll R,
\eeq
where now $\Sigma$ must itself be defined as a
height-integrated, azimuthally and radially averaged quantity.
Mass and angular momentum conservation 
then follow from our equations (\ref{mass})
and (\ref{angmom}):
\beq\label{avgmass}
{\dd\Sigma \over \dd t} + {1\over R} {\dd (R \Sigma \langle
u_R\rangle)\over \dd R} = 0
\eeq
\beq\label{ab}
R^2 \Omega {\dd\Sigma \over \dd t} +
{1\over R}{\dd \over\dd R}\left( R^3\Omega \Sigma\langle
u_R\rangle
+R^2\Sigma W_{R\phi}\right) = 0
\eeq
where $\Sigma W_{R\phi}$ is obtained from
$T_{R\phi}$ after height integration and averaging.
Since $W_{R\phi}$
has dimensions of velocity squared, it is convenient to normalize
this quantity to the isothermal sound speed $c_S^2 =P/\rho$.
Depending on the problem of interest, the value of
$c_S^2$ chosen for normalization might be local, or it could be
a suitably defined global average. 
The ratio $\alpha$ is then a measure of the strength of
the turbulent angular momentum transport (Shakura \& Sunyaev 1973):
\beq
\alpha  c_S^2 \equiv W_{R\phi}.
\eeq
For MHD turbulence, $\alpha$ is an indication of the amplitude
of rms turbulent fluctuations, but this need not be generally true:
it depends upon the existence of a good positive correlation between
the $R$ and $\phi$ components of the stress tensor.  In analytic
modeling $\alpha$ is generally assumed to be constant, but there is
no compelling justification for this, and numerical simulations
show a highly complex spatio-temporal structure for $\alpha$
\citep{np03}.  

Equations (\ref{avgmass}) and (\ref{ab}) lead to an evolutionary
relation between $\Sigma$ and $W_{R\phi}$ \citep{bp99}:
\beq\label{turbevo}
{\dd\Sigma\over \dd t} = {1\over R} {\dd\ \over \dd R}
\left[ {1\over
(R^2\Omega)'} {\dd\ \over \dd R} (\Sigma R^2 W_{R\phi})
\right], 
\eeq
where the prime $'$ indicates $d/dR$.  Diffusive evolution is therefore
not a unique consequence of adopting a viscous stress model.  It follows
instead from mass and angular momentum conservation in a Rayleigh-stable
disk, and the notion of coarse grain averaging implicit in the assumption
that $W_{R\phi}$ exists.

\subsubsection {Thin Disk Energetics}

Consideration of the disk energetics comprises the other
key step in
recovering viscous disk theory.  The dominant terms in the energy 
flux appear to be
\beq\label{eflux} 
( {1\over2} R^2 \Omega^2 +\Phi) \Sigma \langle
u_R\rangle  + \Sigma R\Omega W_{R\phi}.
\eeq
This places a powerful constraint upon thin disk models, since
energy transport introduces no new turbulent correlations into the
equations of motion beyond $W_{R\phi}$.  The point is that $\langle
u_R\rangle$ and $W_{R\phi}$ are tightly restricted by the demands of
mass and angular momentum conservation.  In particular, one is not at
liberty to demand that the total mechanical energy of the fluctuations
plus ``background disk'' be conserved.  Balbus \& Papaloizou (1999) show for
both evolving and steady disks that the mechanical energy loss is in fact
\beq
Q_e = -\Sigma W_{R\phi} {d\Omega\over d\ln R}. 
\eeq
This expression represents true radiative energy losses in a thin disk.
Note, however, that the right hand side is the rate at which free energy
is extracted from the large-scale differential rotation, and is not,
itself, a dissipative quantity (see \S 3.3).  Self-consistent thin disk
theory imposes the very stringent requirement that {\em all}
of this energy be radiated locally.  Non-radiative flows, which are
characterized by $c_S \sim R\Omega$, behave differently. 




\subsection{Transport in Low Radiation Accretion Flows}

Accretion sources in nature show far more variety than the rather tightly
constrained Keplerian thin disk model is capable of, and theorists
have for some time been exploring the behavior of accretion in which
radiative losses play a subdominant role (e.g., Ichimaru 1977; Begelman
1978; Abramowicz et al. 1988).  Interest in these solutions rapidly grew
in the mid-90's when a series of seminal papers appeared (Narayan \&
Yi 1994, 1995) with the goals of constructing simple one-dimensional
models of these accretion profiles and elucidating their signature
observational properties (Narayan, Mahadevan, \& Quataert 1998 for
a review).  The recent unambiguous Chandra detection of the highly
subluminous Galactic Center source Sgr A$^*$ in X-rays (Baganoff et al.\
2001) furnishes compelling evidence that ``dark accretion'' exists in
some guise, and poses a well-defined theoretical challenge (Narayan 2002).

Unlike the case of radiative Keplerian accretion disks, in which there
is at least agreement on the basics of the behavior of simple models,
there is no broad consensus on the dynamical properties of low radiation
accretion flows.  The Sgr A$^*$ detection has made one initially
controversial suggestion --- that electrons in such flows are cooler
than the ions (Narayan \& Yi 1995) --- increasingly hard to avoid.
Yet the following questions continued to be argued in the literature: Are
these flows marked predominantly by inward or outward energy transport?
Are they quasi-spherical?  Do their rotation profiles differ markedly
from Keplerian?  Can they can be dominated by thermal convection?  It has
even been suggested that in the presence of thermal convection, the
turbulent rotating accretion flow is characterized by a vanishing stress
tensor $T_{R\phi}$ and by vanishing dissipation (Narayan, Igumenschev, \&
Abramowicz 2000; Abramowicz et al.\ 2002), though these arguments have
been criticized on both energetic and thermodynamic grounds (Balbus \&
Hawley 2002; Narayan et al.\ 2002 for a response).

Steady bulk accretion is a far from obvious outcome in a gaseous system
which is marginally bound to start with, and subsequently stirred by
the free energy of differential rotation \citep{pr72}, a point that
has recently been visited anew (Blandford \& Begelman 1999).  As in thin
disks, nonradiating gas gradually loses rotational support, but unlike the
radiative models, here it retains a significant level of thermal support.
In fact, even in the absence of any rotation, spherical (Bondi) accretion
is not possible if the adiabatic index exceeds $5/3$, as it effectively
does in a heated, nonradiative gas.  It is thus hardly surprising that in
numerical MHD simulations most of the mass is lost not down the hole,
but through the outer boundaries of the computational grid (Hawley \&
Balbus 2002).

A local viscous prescription for angular momentum transport, which as we
have seen is well-motivated for a rotationally thin disk (see \S 5.6.2),
breaks down completely for a non-radiative flow.  This is a nuisance
for the theorist, but it is inescapable.  Not only is there no longer a
clean separation of turbulent and global scales, there are other turbulent
correlations that appear in the transport equations which are not part
of the $T_{R\phi}$ stress tensor.  The most important of these is the
correlation in velocity and temperature fluctuations seen in equation
(\ref{avgenflux}).  These have no counterparts in {\em viscous} energy
transport, but are in fact much more closely analogous to the wave energy
fluxes of \S 3.

The MRI is the fundamental instability that converts differential
rotation into turbulent fluid motions, whether the system is radiative
or not.  Inward accretion does lead to the development of adverse
entropy gradients however (Narayan \& Yi 1994), which are often then
credited with giving rise to a ``second'' instability.  In fact, the
generalized H\o iland criteria (e.g. eqs.~[\ref{hoil1}], [\ref{hoil2}])
show that these are really secondary features of the same instability.
(Note that in the presence of thermal conduction, an outwardly decreasing
temperature profile is likely to be destabilizing.)  But no secondary
adverse thermal gradient can cause the stress tensor $T_{R\phi}$
to vanish or to change sign in an accretion flow driven by the MRI:
a positive stress is intimately linked to the flow of free energy from
the differential rotation to the turbulent fluctuations.

Despite the enormous output of analytical and quasi-analytical treatments
in the literature, the global behavior of nonradiative, differentially
rotating accretion flows is almost certainly too complex to capture by
standard viscous modeling.  Numerical simulations hold some promise of
telling us, if not everything, then at least where the matter goes.

\section {Numerical Simulations of MHD Disk Turbulence}

\subsection{The Local Approximation}

\subsubsection{Equations and Boundary Conditions}

While attempts are occasionally made at phenomenological theories of
turbulent flow, numerical study has proven to be extremely fruitful.
Both analytic and numerical approaches must each make compromises,
and remain limited in scope.  The principal advantage of numerics is
that brute strength is useful!  The increase in raw computational
power in the last decade has been phenomenal, and simulations of true
three-dimensional turbulence are now a reality.

In the study of accretion disks, two computational schemes are
generally used, local and global.  Global simulations, which are
more difficult, are discussed in section 7. 
In the {\em local approximation}, the idea is to focus on a small region
of the disk, forcing the new origin to corotate with disk fluid orbiting
at a particular
angular velocity, $\Omega_0$.  All velocities are measured relative
to $R\Omega_0$.  In the local limit, curvature terms are ignored.
More formally, the asymptotic regime is defined by
\beq
R\rightarrow \infty, \quad v_\phi \rightarrow \infty, \quad 
{v_\phi\over R} = \Omega\rightarrow
{\rm finite.}
\eeq
We define the velocity $\vecc{w}$:
\beq
\vecc{w} = \vecc{v} - R\Omega_0\, \vvec{e_\phi}.
\eeq
The value of $R$ at which $\Omega = \Omega_0$ will be denoted $R_0$.
In general we shall consider only small radial excursions 
from $R_0$,
\beq
R = R_0 + X, \quad  X \ll R_0.
\eeq
Thus,
\beq
{R} \Omega_0^2 - {\dd\Phi\over \dd R} = R (\Omega_0^2 - \Omega^2(R))\simeq
- \left. X {d\Omega^2\over d\ln R}\right|_{R=R_0}
\eeq
to leading order.  Local Cartesian coordinates can be defined by
aligning the $X$ and $Y$ axes along $R$ and $\phi$.
In a frame rotating at $\Omega$ (dropping the subscript),
the local equations of motion for $w_X$ and $w_Y$ are 
\beq\label{start}
{\dd w_X\over \dd t} + \vvec{w}\cdot\del{w_X} -2\Omega w_Y  + X
{d\Omega^2\over d\ln R} = -{1\over\rho}{\dd P_{tot} \over \dd X}
+ {\vvec{B}\over 4\pi} \cdot\del B_X
\eeq
\beq\label {startbis}
{\dd w_Y\over \dd t} + \vvec{w}\cdot\del{w_Y}
+2\Omega w_X = -{1\over\rho}{\dd P_{tot} \over \dd Y}
+ {\vvec{B}\over 4\pi}\cdot\del B_Y
\eeq
where $P_{tot}$ is the gas plus magnetic pressure.  The forms of the
remaining dynamical equations remain unaffected by the change to rotating
coordinates.

Extended three-dimensional simulations require what are known as
shearing-box boundary conditions.  In this approach, the azimuthal
boundary conditions are always periodic.  Vertical boundary conditions
may be either periodic or outflowing, depending upon the problem
of interest.  Because of the presence of large-scale shear, the radial
boundary conditions require more care to specify (Hawley, Gammie, \&
Balbus 1995).  They are quasi-periodic:  the computational domain is
thought of as one brick in a brick wall extending to infinity, each brick
with the same internal fluid configuration as the next.  The velocity
shear is uniform across the entire brick wall, so one layer of bricks
slides with respect to the layer above and below!  A fluid element
leaving a radial boundary of the one-brick computational domain
is replaced by its image on the opposite side, but not, in general,
at the azimuth it has just vacated.  Rather, the image reappears at the
azimuth of the sliding brick that the original fluid element is entering.
The mathematical formulation of these boundary conditions may be found
in Balbus \& Hawley (1998).

\subsubsection {Local Two-Dimensional Simulations}

In the first extended axisymmetric simulations carried out by Hawley \&
Balbus (1992), a surprise emerged.  Though the evolution of an initial
uniform vertical magnetic field was in accord with analytic theory in the
linear stages of its development, the nonlinear stages did not appear
to be turbulent at all.  In fact, the linear stage seemed to continue
indefinitely, with exponentially growing streaming motions persisting.
This contrasted sharply with the nonlinear behavior of a shearing
box starting with a vertical field whose mean value was zero---half
upwards and half downwards say, or sinusoidally varying.  In that case,
turbulence quickly developed after a few linear growth times, and then
gradually decayed, leaving no field at all at the end of the simulation.

The latter behavior is a consequence of Cowlings' anti-dynamo theorem
(e.g. Moffatt 1978), which prohibits field amplification in an isolated
axisymmetric MHD flow.  The case of a mean vertical field avoids this
restriction because it does not meet the requirement of isolation:
the field extends to infinity.  This is not necessarily unphysical:
real disks certainly can be threaded by a magnetic field which closes
at very large distances away from the midplane.  The question is, do such
disks exhibit the streaming behavior discussed above in three-dimensions?


\subsubsection{Local Three-Dimensional Simulations}

The study of true dynamo amplification requires the implementation
of three-dimensional MHD codes.  The simplest approach consists of a
homogeneous shearing box, in which only the radial component of the
large-scale gravitational field is retained (Hawley et al. 1995; Matsumoto
\& Tajima 1995).  The initial field geometry in these studies had some
combination of vertical and toroidal components.  A more complicated
initial field configuration, important for investigating dynamo activity,
is to give the initial field a random character with vanishing mean value
(Hawley, Gammie, \& Balbus 1996).  Finally, adding a vertical component of
the gravitational field produces a density stratification (Brandenburg et
al.\ 1995; Stone et al.\ 1996), and introduces buoyancy into the problem.

The answer to the question posed at the end of the previous subsection
is that three-dimensional studies in fact do show the breakdown of
the two-dimensional streaming solutions (Hawley et al.~1995).  If the
computational box is large enough to allow an unstable radial wavelength,
streaming is disrupted within a few orbits, as the Kelvin-Helmholtz
instability noted by Goodman \& Xu (1994) leads to fluid turbulence,
and of course its large associated angular momentum transport.

Later work by Stone et al.~(1996) introduced vertical stratification,
allowing for two vertical scale-heights in an initially isothermal disk.
With the onset of the MRI, the magnetic field emerged from the disk to
establish a highly magnetized corona.  However, the amplitude of the
turbulence near the midplane is determined more by local dissipation than
by buoyant losses, and the resulting stress and transport levels are not
greatly modified from the nonstratified simulations.  The presence of
a corona, however, may well have important observational consequences,
especially if a significant amount of dissipational heating occurs there.
Miller \& Stone (2000) performed simulations with a larger computational
domain covering 5 scale heights above and below the equator.  They showed
that a strongly magnetized corona ($P_{\rm gas} \ll P_{\rm mag}$) forms
rather naturally from little more than a disk and a weak seed field.

\subsection{Radiative Effects}

The radiation energy density in the inner regions of black hole accretion
rivals or exceeds the thermal energy density.  The linear stability
of a magnetized, stratified, radiative gas was recently addressed by
Blaes \& Socrates (2001).  Despite the complexity of the full problem,
the MRI emerges at the end of the day unscathed, its classical stability
criterion $d\Omega^2/dR > 0$ remaining intact.  The maximum growth rate
can be lowered, however, when the azimuthal field approaches or exceeds
thermal strength.  (The same is true for the ordinary MRI, as shown by
Blaes \& Balbus [1994].)

The proper interpretation of a linear analysis is slightly unclear,
since the unperturbed state is presumably fully turbulent due to the MRI;
a rotationally stable disk lacks an internal energy source.  The interplay
between MHD turbulence and radiation is intrinsically nonlinear.  Turner,
Stone, \& Sano (2002), have studied a local, radiative, axisymmetric,
shearing box.  The linear calculations of Blaes \& Socrates (2001)
were confirmed in detail, and the nonlinear flow fully developed.
As in standard MRI simulations, the stress is dominated by the Maxwell
component, which is a factor of a few larger than the Reynolds terms.
Photon diffusion maintains nearly isothermal conditions while creating
over-dense small clumps of thermally-supported gas.  This occurs when
radiation pressure support is rapidly lost via small wavelength diffusion.
Optically thick radiative disks may therefore be more inhomogeneous than
their gaseous counterparts.

\subsection{Low Ionization Disk Simulations}  Protostellar disks, and
possibly cataclysmic variable disks, contain regions of low ionization
fraction---so low that the assumptions of ideal MHD break down.  Ohmic
dissipation becomes important, together with the Hall inductive terms.
The effects of Hall electromotive forces on the MRI have been studied
by Wardle (1999) and by Balbus \& Terquem (2001).  Nonlinear numerical
simulations including ohmic resistivity have been done by Fleming, Stone,
\& Hawley (2000); Sano \& Stone (2002) have done simulations including
both the ohmic and Hall processes.  Under rather general conditions,
if Ohmic dissipation is important, Hall electromotive forces are also
important (Balbus \& Terquem 2001; Sano \& Stone 2002).

The most interesting feature introduced by Hall electromotive forces is
helicity: the relative orientation of the angular velocity
$\vvec{\Omega}$ and magnetic field $\vvec{B}$ vectors matters (Wardle
1999; Balbus \& Terquem 2001).  The key point is that
$\vvec{\Omega}\bcdot \vvec{B} > 0 $ configurations raise the maximum
growth rate in the presence of ohmic losses, and this aligned
configuration results in more vigorous transport in the local
axisymmetric simulations of Sano \& Stone (2002).  In configurations
where $\vvec{\Omega}\bcdot \vvec{B}$ vanishes on average, the level of
turbulence is much more sensitive to the size of the ohmic dissipation
term.

In the simulations of Fleming et al.~(2000), a critical magnetic Reynolds
number $Re_M$ emerged below which turbulence is suppressed.  ($Re_M$ is
defined here as the ratio of the product of the scale height and sound
speed to the resistivity.)  When the mean field vanishes, $Re_M$ was
found to be surprisingly high, $\sim 10^4$.  In the presence of a mean
field, the critical $Re_M\sim 10^2$.  The interesting and important
question is whether the inclusion of Hall electromotive forces changes
these numbers by lowering them, i.e., making it easier to support
turbulence.  The Sano \& Stone (2002) axisymmetric simulations did not
reveal a large change, but questions on the maintenance of turbulence
more properly await a three-dimensional treatment.

\section{Global Disk Simulations}

Global accretion simulations are a demanding computational problem,
requiring extended evolution at high resolution.  Resolving the
turbulent cascade through its full inertial range is not yet possible.
The global disk structure alone extends from the black hole horizon,
$r_S$, out to $\sim 10^3$ $r_S$.  Time scales $\propto \Omega$, so that
Keplerian rotation implies a factor of $\sim 3\times 10^4$ in time scales.
While compromises are unavoidable, the problems of interest are so rich
that much can be learned even from highly idealized flows.

\subsection{Two-Dimensional Simulations}

Given the difficulties of these simulations, it is somewhat surprising
to note that global axisymmetric MHD disk studies have been around
for almost two decades, largely motivated by jet formation problems
(Uchida \& Shibata 1985).  It is only more recently that two-dimensional
simulations have focused on the internal dynamics of the disk
itself and the resulting accretion flow, rather than on the launching
and collimation of jets.


The challenge of understanding under-luminous accreting compact objects
has been, from the numericist's perspective, a most welcome development.
This is precisely the type of flow that is well-suited to numerical
simulation.  The first two-dimensional MHD simulations of nonradiating
accretion flows were carried out by Stone \& Pringle (2001).  In this
study, accretion begins from a constant angular momentum torus, with weak
embedded poloidal field loops.  The initial infall phase is relatively
smooth, and marked by streaming.  Subsequent evolution is decidedly
turbulent.  The resulting flow consists of an approximately barotropic
disk near the midplane, with constant $\Omega$ contours parallel to
cylindrical radii.  A substantial magnetized coronal outflow enveloped the
disk.  Finally, there was very little difference in the character of the
solution if dissipative losses were retained as heat or simply ignored.

Toward the end of the Stone \& Pringle (2001) simulations, the turbulence
is noticeably waning, because there is nothing to save it from the
anti-dynamo theorem.  Axisymmetric simulations also tend to over-emphasize
radial streaming, and cannot allow for toroidal field instabilities.
Nothwithstanding these limitations,  by allowing a rapid investigative
turnover of plausible accretion scenarios, two-dimensional simulations
have proven to be a valuable tool, and a remarkably reliable one as well.

\subsection{Cylindrical Disks}

The ``cylindrical disk'' is a global three-dimensional system allowing for
full radial and azimuthal dynamics, but ignoring the vertical component
of the central gravitational field.  Although there is no vertical
stratification, turbulence and magnetic fields can be sustained with far
fewer vertical grid zones than required for a full three-dimensional
simulation.

Armitage (1998) carried out the first such calculation using the standard
ZEUS code and a grid covering the full $2\pi$ in azimuth, running from
$R=1$ to 4, and $Z=0$ to $0.8$. The radial boundaries were reflecting,
the vertical boundaries periodic.  The initially vertical magnetic
field was taken proportional to $\sin(kR)/R$, with the sine function
argument varying over $2\pi$ between $R=1.5$ and 3.5.  MHD turbulence
rapidly developed.

A more extensive set of cylindrical simulations (Hawley 2001) reaffirmed
many of the conclusions of the local box models, and tested several
technical aspects of global simulations.  Restricting the azimuthal
range to some fraction of $2\pi$ does not seem to affect the qualitative
outcome of a simulation, nor does the polytropic index in the equation
of state.  Simulations that start with a vertical field, however,  have
greater amplification and higher ratios of stress to magnetic pressure
compared with those beginning with toroidal fields.

Cylindrical disk calculations have been applied to the problem of the
star-disk boundary layer (Armitage 2002; Steinacker \& Papaloizou 2002).
Such simulations clearly represent an important step forward beyond
viscous models.  Both sets of investigators reported dynamo amplification.
In particular, Armitage (2002) finds an order of magnitude larger field
energy density in the boundary layer compared with the average disk field.
Dissipative heating in the boundary layer has yet to be simulated.

\subsection{Three-Dimensional Simulations}

Global three-dimensional MHD disk simulations with complete vertical
structure have been discussed by Hawley (2000).   Here, the initial
configuration is a constant angular momentum tori with a weak
embedded magnetic field, which could be either toroidal or poloidal.
Such structures appear in models of active galactic nuclei, where it
is thought that they feed an inner disk (Krolik 1999).  But the most
notable feature of Hawley's simulations is the speed with which an initial
constant angular momentum profile is changed to nearly Keplerian---a few
orbital times at the pressure maximum of the initial torus.  The weak
field instability is so efficient at transporting angular momentum,
that vigorous radial spreading dilutes the structural pressure gradient,
leaving the disk rotationally supported.  At later times in the course
of the simulation, the disks show strong time and space variability
on all resolvable scales, and tightly wrapped spiral structure.
There is significant angular momentum transport: typical $\alpha$
values are several percent, with pronounced and rapid fluctuations.
Similar results were obtained (for toroidal fields) by Machida, Hayashi,
\& Matsumoto (2000).

Steinacker \& Henning (2001) revisit the question of the influence of a
large-scale vertical field on an accretion disk, previously studied in
axisymmetry.  They obtain very high accretion rates, which is described
as a ``collapse.''  Coupling of the disk to a magnetized corona also
appears to drive outflows, with the degree of collimation depending on
the strength of the field.

Hawley \& Krolik (2001, 2002) investigate the evolution of a
magnetized accretion torus lying near the marginally stable orbit in
a pseudo-Newtonian model potential for the black hole (Paczy\'nsky \&
Wiita 1980).  These papers focus on the behavior of the stress tensor
in gas passing through the so-called plunging region inside of the
marginally stable circular orbit.  In contrast to standard viscous models,
the simulations suggest that the disk does not sharply truncate at the
marginally stable orbit, nor does the stress vanish here.  (The stress
actually increases somewhat in the plunging region, owing to the strong
correlation engendered in the radial and azimuthal field components.)
Once again, variability is seen over a large range of spatial and time
scales.

\begin{figure}[t]
\centerline {\epsfbox{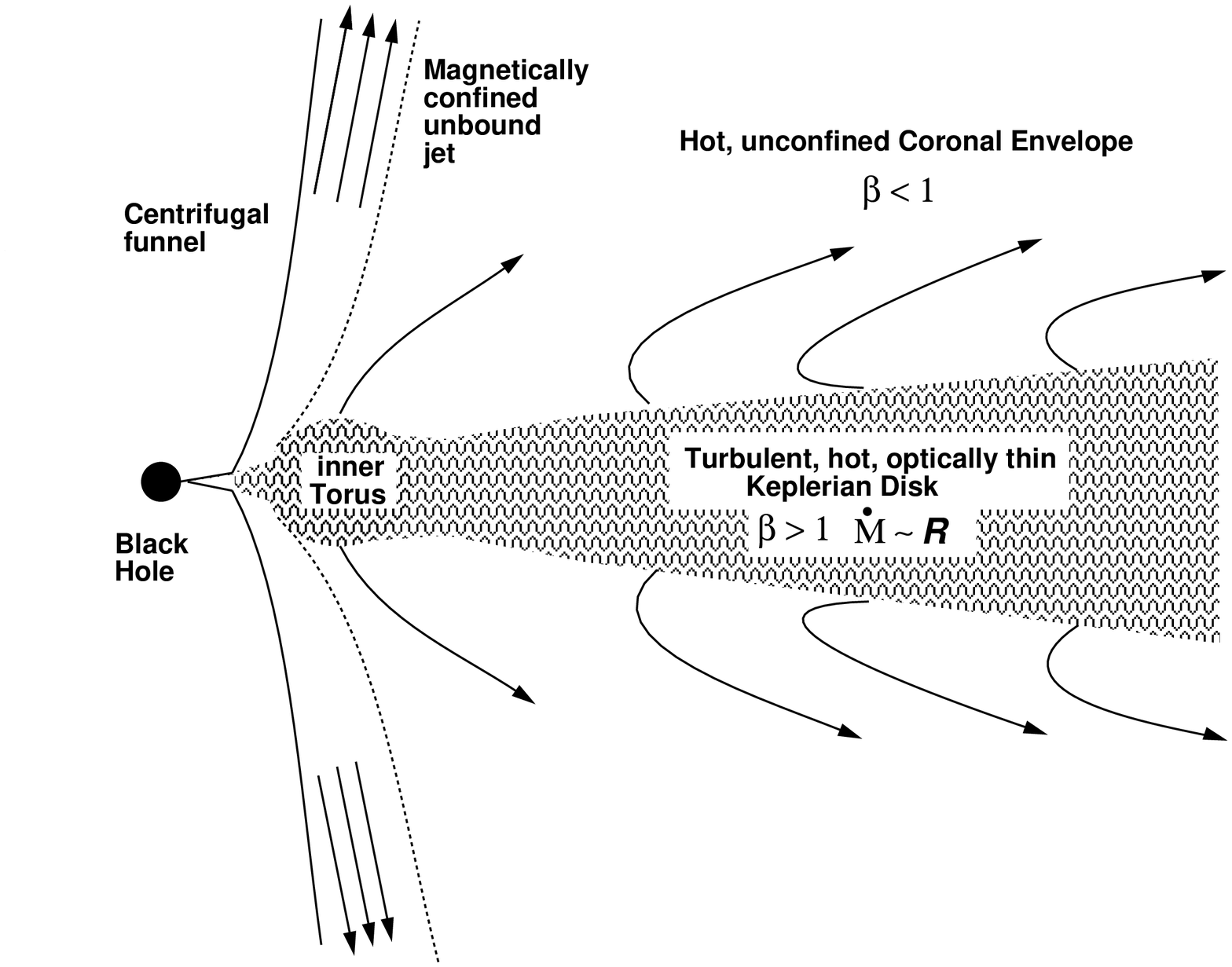}\hfill}
\caption{
A schematic diagram of a nonradiative accretion flow, highlighting its
principal features.  A turbulent, nearly Keplerian gas-dominated hot
disk is surrounded by an active, diffuse, magnetic-dominated coronal
envelope.  Near the marginally stable orbit, the flow thickens into a
small inner torus.  A centrifugally-evacuated funnel lies along the
axis, surrounded by a jet confined by magnetic pressure in the corona.
From Hawley \& Balbus (2002).} \label{eps1} \end{figure}
 
The more general problem of nonradiative global accretion over larger
radial domains, first considered in two dimensions by Stone \& Pringle
(2001), has been extended to three dimensions by Hawley, Balbus \&
Stone (2001) and Hawley \& Balbus (2002).  Once again the accretion
originates with a constant specific angular momentum torus.  But in
these simulations, the torus is initially centered out rather far out,
at 100 Schwarzschild radii from the hole.   The resulting turbulent flow
appears at the end of the simulation to settle into three well-defined
dynamical structures.  What accretion is present is via a hot, thick,
slightly sub-Keplerian disk, which evidently is characteristic of
radiative and nonradiative disks alike.  Surrounding this disk is a
strongly magnetized corona with vigorous circulation and perhaps a
weak outflow.  Magnetic coronal pressure confines a jet-like outflow,
pinning it against the centrifugal funnel wall, an effective equipotential
surface.

Figure (\ref{eps1}) shows schematically the appearance of the disk,
corona, and jet as they appear in a typical run.  The inner torus is
pressure-thickened, but remains predominately supported by rotation.
It is a transitory structure, forming and collapsing over the
course of the simulation.

%
%
%

\section{Summary}

Our understanding of the basis of astrophysical accretion has deepened
enormously in the past decade, and the physical process underlying
the ``anomalous viscosity'' of accretion disks has been elucidated.
I believe that the term itself should be avoided, since it is so
misleading.  Magnetic fields do much more than fill the role occupied
by a hydrodynamical Navier-Stokes viscosity.  The behavior of magnetized
fluids is far too subtle for this approach to be profitably carried into
its fourth decade.

A one sentence summary of the effects of weak magnetic fields on the
stability of a gaseous system is as follows:  Magnetic fields turn 
gradient free energy sources (angular velocity and temperature) into
sources of dynamical instability.  Provided the assumptions inherent in
the MHD approximation remain valid, the fields maintain this influence
even in the formal limit $B\rightarrow 0$.   This is why any formal
stability analysis on an {\em ersatz} viscous gas is likely to be
misleading, at best.

The combination of a magnetic field and outwardly decreasing differential
rotation is thus prone to what is generally known as the magnetorotational
instability, or MRI for short.  Localized simulations of Keplerian disks
indicate that the MRI leads to a turbulence-enhanced stress tensor that
transports energy and angular momentum outwards, allowing accretion to
proceed.  The typical dimensionless value of the stress (normalized to
a fiducial pressure) ranges from $5\times 10^{-3}$ to $0.6$ depending
upon field geometry, and is highly variable in both space and time.
Local simulations have become rather sophisticated in the range of
problems they are able to address, and work has now begun to include
radiation dynamics and non-ideal MHD.   There is still considerable
uncertainty of how magnetic fields behave in disks containing regions
of low ionization, and hydrodynamical processes certainly cannot be
ruled out.  Wave transport, of interest in its own right as well as for
the perspective it brings to turbulent transport, is likely to be central
to our understanding of protoplanetary disk evolution.   Studies of the
interaction between density waves and MHD turbulence have barely begun.

Global three-dimensional MHD simulations are now being run by several
groups of investigators.  The most robust finding of these studies is
the speed with which a Keplerian profile emerges from an initial thick
torus configuration, whatever the field geometry.  The most detailed
simulation performed to date (Hawley \& Balbus 2002) ends in a three
component structure: a warm Keplerian disk, a highly magnetized corona,
and an axial jet.  Considerably more exploration needs to be done before
we can be certain of whether this outcome is generic.

The unique spatial resolution and sensitivity capabilities of the
Chandra X-ray satellite provide an auspicious opportunity to bridge
the gap between computational gas dynamics and observed spectra.
Chandra observations of the Galactic Center provide compelling evidence
that very low luminosity accretion flows are present in nature, and
these the numericist may hope to simulate with perhaps some fidelity.
Relatively little has been done to elicit the radiative properties of
the vast digital accretion flow data base.  There are grounds for some
hope that this untapped resource may advance our understanding of black
hole accretion as much as has the last remarkable decade.


\section*{Acknowledgements} It is a pleasure to thank  S.\ Fromang,
J.\ Hawley , and K.\ Menou for reading an early version of this review.
All made constructive comments that greatly improved the presentation.
Support under NSF grant AST-0070979, and NASA grants NAG5-9266 and
NAG-10655 is gratefully acknowledged.


\begin{thebibliography}{}

\bibitem[Abramowicz et al. 1988] {aet88} Abramowicz, M.~A., Czerny, B.,
Lasota, J.~P., \& Szuszkiewicz, E. 1988.\ \apj 332:646--58

\bibitem[Abramowicz, et al.\ (2002)]{aiqn02} Abramowicz, M.~A.,
Igumenschev, I.~V., Quataert, E., \& Narayan, R. 2002, ApJ, 565,
1101

\bibitem[Acheson \& Gibbons 1978]{ag78} Acheson, D.~J., \& Gibbons, M.~P.\
1978.\ {\it Philosophical Transactions of the Royal Society of
Londn.\ Series A, Mathematical and Physical Sciences.} 289,
1363:459--500

\bibitem[Acheson \& Hide 1972]{ah72} Acheson, D.~J., \& Hide, R. 1972.\ 
{\it Rep.\ Prog.\ Phys.} 36:159--221

\bibitem[Armitage 1998]{a98} Armitage, P.~J. 1998.\ \apj 501:L189--92

\bibitem[Armitage 2002]{a02} Armitage, P.~J. 2002.\ \mn 330:895--900

\bibitem[Baganoff, et al.~(2001)]{bag01} Baganoff, F.~K., Bautz, M.~W.,
Brandt, W.~N., Chartas, G., Feigelson, E.~D., Garmire, G., Maeda, Y.,
Morris, M., Ricker, G.~R., Townsley, L.~K., \& Walter, F.\  2001, Nature
413: 45-8

\bibitem[Balbus 1995]{b95} Balbus, S.~A.\ 1995.\ \apj 453:380--3

\bibitem[Balbus 2000]{b00} Balbus, S.~A.\ 2000.\ \apj 534:420--27

\bibitem[Balbus 2001]{b01} Balbus, S.~A.\ 2001.\ \apj 562:909-17

\bibitem[Balbus 2002]{b02} Balbus, S.~A.\ 2002.\ In {\em The Physics
of Cataclysmic Variables and Related Objects.} eds.\ B.~T.\
G\"ansicke, K.\ Beuermann, \& K.\ Reinsch.\ A.S.P., San Francisco.\
pp.\ 356--66

\bibitem[Balbus \& Hawley 1991]{bh91} Balbus, S.~A., \& Hawley,
J.~F.\ 1991.\ \apj 376:214--22

\bibitem[Balbus \& Hawley 1992]{bh92} Balbus, S.~A., \& Hawley,
J.~F.\ 1992.\ \apj 392:662--66

\bibitem[Balbus \& Hawley 1998]{bh98} Balbus, S.~A., \& Hawley, J.~ F.\
1998.\ {\it Rev.\ Mod.\ Phys.} 70:1--53

\bibitem[Balbus \& Hawley 2002]{bh02} Balbus, S.~A., \& Hawley,
J.~F.\ 2002.\ \apj 573:749--53

\bibitem[Balbus \& Terquem 2001]{bt01} Balbus, S.~A., \& Terquem, C.\
2001.\ \apj 552:235--47

\bibitem[Balbus, Hawley, & Stone 1996]{bhs96} Balbus, S.~A., Hawley,
J.~F., \& Stone J.~M. 1996. \ApJ 467:76--86

\bibitem[Balbus \& Papaloizou 1999]{bp99} Balbus, S.~A., \&
Papaloizou, J.~C.~B.\ 1999.\ {\apj}521:650--58

\bibitem[Bayly et al.\ 1988]{betal88} Bayly, B.~J., Orszag, S.~A., \&
Herbert, T. 1988.\ {\it Ann.\ Rev.\ Fluid Mech.}20:359--91

\bibitem[Begelman 1978]{b78} Begelman, M.\ 1978.\ \mn 184:53--67

\bibitem [Binney \& Tremaine 1987]{bt87} Binney, J., \& Tremaine, S.
1987, {\it Galactic Dynamics,} Princeton Univ.\ Press, Princeton

\bibitem[Blaes 1987]{b87} Blaes, O.~M.\ 1987.\ \mn 227:975--92

\bibitem[Blaes 2002]{bl02} Blaes, O.~M.\ 2002.\ {\it Physics Fundamentals
of Luminous Accretion Disks Around Black Holes,} in ``Accretion Disks,
Jets, and High Energy Phenomena in Astrophysics'', Proceedings of Session
LXXVIII of Les Houches Summer School, Chamonix, France, August 2002,
eds.\ F.\ Menard, G.\ Pelletier, G.\ Henri, V.\ Beskin, and J.\ Dalibard \
EDP Science: Paris and Springer, Berlin, in press (astro-ph/0211368)

\bibitem[Blaes \& Balbus 1994]{bb94} Blaes, O.~M., \& Balbus, S.~A.\
1994.\ \apj 421:163--77

\bibitem[Blaes \& Socrates 2001]{bs01} Blaes, O.~M., \& Scorates,
A.\ 2001.\ \apj 553:987--98

\bibitem[Blandford \& Begelman 1999]{bb99} Blandford, R.~D., \& Begelman, 
M.~C. 1999.\ \mn 303:L1--5

\bibitem[Brandenburg et al.~(1995)]{betal95} Brandenburg, A., Nordlund,
\AA., Stein, R.~F., \& Torkelsson, U.\ 1995.\ \ApJ 446:741--54

\bibitem[Cabot 1996]{c96} Cabot, W.\ 1996.\ \apj 465:874--86

\bibitem[Chandrasekhar 1953]{c53} Chandrasekhar, S.\ 1953. {\it Proc.\
Roy.\ Soc.\ (London) A} 216:293--309

\bibitem[Chandrasekhar 1960]{c60} Chandrasekhar, S.\ 1960. {\it Proc.\
Nat.\ Acad.\  Sci.\ }  46:253--7  

\bibitem[Chandrasekhar 1961]{c61} Chandrasekhar, S.\ 1961. {\it
Hydrodynamic and Hydromagnetic Instability,} Clarendon Press, Oxford,\
pp.\ 402--3

\bibitem[Coles 1965]{c65} Coles, D. 1965.\ {\it J.\ Fluid Mech.\ }
21:385--425

\bibitem[Corcos \& Lin 1984]{cl84} Corcos, G.~M., \& Lin, S.~J.\ 1984.\
{it J.\ Fluid Mech.\ } 139:67--95

\bibitem[Crawford \& Kraft 1956]{ck56} Crawford, J.~A., \& Kraft,
R.~P.\ 1956.\ \apj 123:44--53

\bibitem[Fleming, Stone, \& Hawley 2000]{fsh00} Fleming, T.~P.,
Stone, J.~M., \& Hawley, J.~F.\ 2000.\ \apj 530:464--77

\bibitem[Frank, King, \& Raine 2002]{fkr02} Frank, J., King, A., 
\& Raine, D.\ 2002.\ {\it Accretion Power in Astrophysics,}
Cambridge Univ. Press, Cambridge, pp.\ 104--5

\bibitem[Fricke 1969]{f69} Fricke, K.\ 1969.\ \asap 1:388--98

\bibitem[Fromang, Terquem, \& Balbus 2002]{ftb02} Fromang, S., Terquem,
C., \& Balbus, S.~A.\ 2002 \mn 329:18--28

\bibitem[Gammie 1996]{g96} Gammie, C.~F.\ 1996.\ \apj 457:355--62

\bibitem[Gammie \& Menou 1998]{gm98} Gammie, C.~F., \& Menou, K.\
1998.\ \apj 492:L75--8

\bibitem[Glassgold, Feigelson, \& Montmerle 2000]{gfm00} Glassgold, A.~E.,
Feigelson, E.~D., \& Montmerle, T. 2000.\ In {\it Protostars and Planets IV.}
eds. V. Mannings, A.~P. Boss, \& S.~S. Russell.\  Univ.\ Arizona, Tucson.\
pp.\ 429--55

\bibitem[Goldreich et al.~(1986)]{ggn86} Goldreich, P., Goodman, J.,
and Narayan, R. 1986.\ \mn 221:339--64

\bibitem[Goldreich \& Tremaine 1979]{gt79} Goldreich, P., \&
Tremaine, S.~D.\ 1979.\ \apj 233:857--71

\bibitem[Goldreich \& Tremaine 1980]{gt80} Goldreich, P., \&
Tremaine, S.~D.\ 1980.\ \apj 241:425--41 

\bibitem [Goodman \& Ji 2002]{gj02} Goodman, J., \& Ji, H. 2002.\ {\it J.\
Fluid Mech.} 462:365--82

\bibitem[Goodman \& Rafikov 2001]{gr2001} Goodman, J., \& Rafikov,
R.~R. 2001.\ \apj 552:793--802

\bibitem[Goodman \& Xu 1994]{gx94} Goodman, J., \& Xu, G. 1994.\
\apj 432:213--23

\bibitem[Hartmann et al.\ 1998]{hetal98} Hartmann, L., Calvet, N.,
Gullbring, E., \& D'Alessio, P.\ 1998. \apj 495:385--400

\bibitem[Hawley 1991]{h91} Hawley, J.~F.\ 1991.\ \apj 381:496--507

\bibitem[Hawley 2000]{h00} Hawley, J.~F.\ 2000.\ \apj 528:462--79

\bibitem[Hawley 2001]{h01} Hawley, J.~F.\ 2001.\ \apj 554:534--47

\bibitem[Hawley \& Balbus 1992]{hb92} Hawley, J.~F., \& Balbus,
S.~A.\ 1992.\ \apj 400:595--609

\bibitem[Hawley \& Balbus 2002]{hb02} Hawley, J.~F., \& Balbus,
S.~A.\ 2002.\ \apj 573:738--48   

\bibitem[Hawley, Balbus, \& Stone 2001]{hbs01} Hawley, J.~F., Balbus,
S.~A., \& Stone, J.~M.\ 2001.\ \apj 554:L49--52  

\bibitem[Hawley, Balbus, \& Winters (1999)]{hbw99} Hawley, J.~F.,
Balbus, S. A., \& Winters, W. F. 1999.\ \ApJ 518:394--404

\bibitem[Hawley, Gammie \& Balbus (1995)]{hgb95} Hawley, J.~F.,
Gammie, C.~F., \& Balbus, S. A.  1995.\ \ApJ 440:742--63

\bibitem[Hawley, Gammie \& Balbus (1996)]{hgb96} Hawley, J.~F.,
Gammie, C.~F., \& Balbus, S. A.  1996.\ \ApJ 464:690--703

\bibitem[Hawley \& Krolik 2001]{hk01} Hawley, J.~F., \& Krolik,
J.~H.\ 2001.\ \apj 548:348--67

\bibitem[Hawley \& Krolik 2002]{hk02} Hawley, J.~F., \& Krolik,
J.~H.\ 2002.\ \apj 566:164--80

\bibitem[Hawley, \& Stone (1995)]{hs95}Hawley, J.~F., \& Stone, J.~M.
1995.\ {\it Comp.\ Phys.\ Comm.} 89:127--48

\bibitem[Ichimaru (1977)]{i77} Ichimaru, S. 1977.\ \ApJ 214:840--55

\bibitem[Ji, Goodman, \& Kageyama 2001]{jgk01} Ji, H., Goodman, J., \&
Kageyama, A.\ 2001.\ MNRAS, 325, L1--5

\bibitem[Krolik 1999]{k99} Krolik, J.\ 1999.\ {\it Active Galactic
Nuclei,} Princeton University Press, Princeton

\bibitem[Li, et al.\ 2000]{lietal00} Li, H., Finn, J.~M., Lovelace,
R.~V.~E., \& Colgate, S.~A.\ 2000.\ \ApJ 533:1023--34

\bibitem[Lighthill 1978]{l78} Lighthill, J. 1978.\ {\it Waves in
Fluids,} Cambridge: Cambridge Univ.\ Press 

\bibitem[Lin \& Papaloizou 1979]{lp79} Lin, D.~N.~C., \& Papaloizou,
J.~C.~B.\ 1979.\ \mn 188:191--201

\bibitem[Longaretti 2002]{l02} Longaretti, P.-Y. 2002.\ \ApJ 576:587--98

\bibitem[Lovelace et al.~(1999)]{lovetal99} Lovelace, R.~V.~E., Li,
H., Colgate, S.~A., \& Nelson, A.~F.\ 1999.\ \ApJ 513:805--10

\bibitem[Lynden-Bell 1969]{lb69} Lynden-Bell, D.\ 1969.\ {\it Nature}
223:690--94

\bibitem[Lynden-Bell \& Kalnajs 1972]{lbk72} Lynden-Bell, D., \& Kalnajs, A.~J.\
1972.\  \mn 157:1-30

\bibitem[Lynden-Bell \& Pringle 1974]{lbp74} Lynden-Bell, D., \&
Pringle., J.~E.\ 1974. \mn 168:603--37

\bibitem[Machida, Hayashi, \& Matsumoto 2000]{mhm00} Machida, M., 
Hayashi, M.~R., \& Matsumoto, R.\ 2000.\ \apj 532:L67--70

\bibitem[Matsumoto \& Tajima 1995]{mt95} Matsumoto, R., \& Tajima, T.
1995.\ \apj 445:767--79

\bibitem[Menou \& Quataert 2001]{mq01} Menou, K., \& Quataert, E.\
2001.\ \apj 552:204-8

\bibitem[Miller \& Stone 2000]{ms00} Miller, K.~A., \& Stone, J.~M.
2000.\ \apj 534:398--419

\bibitem[Moffatt 1978]{m78} Moffatt, K. 1978, {\it Magnetic Field
Generation in Electrically Conducting Fluids.}  Cambridge University
Press, Cambridge

\bibitem[Narayan 2002]{n02} Narayan, R. 2002.\ In {\it Lighthouses of
the Universe: The Most Luminous Celestial Objects and Their Use for
Cosmology, Proceedings of the MPA/ESO,} eds.\ M. Gilfanov, R.\
Sunyaev, E.\ Churazov, pp.\ 405--29.\ Berlin:  Springer-Verlag 

\bibitem[Narayan, Igumenschev, \& Abramowicz (2000)]{nia00} Narayan,
R., Igumenschev, I.~V., \& Abramowicz, M.~A. 2000, ApJ, 539, 798

\bibitem[Narayan, Mahadevan, \& Quataert 1998]{nmq98} Narayan, R.,
Mahadevan, R., \& Quataert, E. 1998.\ In {\it Theory of Black Hole
Accretion Disks,} eds.\ M.~A.~Abramowicz, G.~Bjornsson, and J.~E. Pringle,
Cambridge University Press, Cambridge.\ pp. 148--73

\bibitem[Narayan et al. 2002]{netal02} Narayan, R., Quataert, E., 
Igumenschev, I.~V., \& Abramowicz, M.~A. 2002.\ \apj 577:295--301

\bibitem[Narayan \& Yi]{ny94} Narayan, R., \& Yi, I. 1994.\ \ApJ
428:L13--16

\bibitem[Narayan \& Yi 1995]{ny95} Narayan, R., \& Yi, I. 1995.\ \ApJ
452:710--35

\bibitem[Nelson \& Papaloizou 2003]{np03} Nelson, R.~P., \& Papaloizou,
J.~C.~B. 2002, astro-ph/0211495

\bibitem[Newcomb 1962]{n62} Newcomb, W.~A.\ 1962. {\it Nuc.\ Fusion:
1962 Suppl., Part 2.}  pp.\ 451--63

\bibitem[Noguchi et al.\ 2002]{ngetal02} Noguchi, K., Pariev, V.~I.,
Colgate, S.~A., Beckely, H.~F., \& Nordhaus, J. 2002.\ \ApJ 575:151--62

\bibitem[Ogilive \& Lubow 1999]{ol99} Ogilvie, G.~I., \& Lubow, S.~H.\
1999.\ \apj 515:767--75

\bibitem[Ogilive \& Proctor 2003]{op03} Ogilvie, G.~I., \& Proctor,
M.R.~E.\ 2003.\ {\it J.\ Fluid Mech.\ } 476:389--409

\bibitem[Orszag \& Patera 1980]{op80} Orszag, S.~A., \& Patera,
A.~T.\ 1980.\ {\it Phys.\ Rev.\ Lett.} 45:989--93

\bibitem[Orszag \& Patera 1981]{op81} Orszag, S.~A., \& Patera,
A.~T.\ 1981. In {\it Transition and Turbulence,} ed.\ R.~E. Meyer,
pp.\ 127--46. New York: Academic

\bibitem[Paczy\'nsky \& Wiita 1980]{pw80} Paczy\'nsky, B., \&
Wiita, P.~J.\ 1980.\ \asap 88:23--31

\bibitem[Papaloizou \& Pringle 1984]{pp84} Papaloizou, J.~C.~B., \&
Pringle, J.~E.\ 1984.\ \mn 208:721--50

\bibitem[Papaloizou \& Pringle 1985]{pp85} Papaloizou, J.~C.~B., \&
Pringle, J.~E.\ 1985.\ \mn 213:799-820

\bibitem[Papaloizou \& Terquem]{pt95} Papaloizou, J.~C.~B., \&
Terquem, C.\ 1995.\ \mn 274:987--1001


\bibitem[Pierrehumbert \& Widnall 1982]{pw82} Pierrehumbert, R.~T.,
\& Widnall, S.~E. 1982.{\it J.\ Fluid Mech.\ } 114:59--82

\bibitem[Pringle \& Rees 1972]{pr72} Pringle, J.~E., \& Rees, M.~J.
1972, AA, 21, 1

\bibitem[Richard \& Zahn 1999]{rz99} Richard, D., \& Zahn J.-P. 1999.\
\asap 347:734--38

\bibitem[R\"udiger \& Shalybkov 2002]{rs02} R\"udiger, G., \&
Shalybkov, D.\ 2002, Phy.\ Rev.\ E 66:016307-(1--8)

\bibitem[Sano \& Stone 2002]{ss02} Sano, T., \& Stone, J.~M. 2002.\
\apj 570:314--28

\bibitem[Shakura \& Sunyaev 1973]{ss73} Shakura, N.~I., \& Sunyaev,
R.~A. 1973.\ \asap 24:337--55

\bibitem[Shu 1992]{shu92} Shu, F.~H.\ 1992.\ {\it The Physics of Astrophysics.
Gas Dynamics.}  University Science, Sausalito

\bibitem[Steinacker \& Henning 2001]{sh01} Steinacker, A., \&
Henning, T.\ 2001.\ \apj 554:514--27

\bibitem[Steinacker \& Papaloizou 2002]{sp02} Steinacker, A., \&
Papaloizou, J.~C.~B.\ 2002.\ \apj 571:413--28

\bibitem[Stone \& Balbus 1996]{sb96} Stone, J.~M., \& Balbus,
S.~A.\ 1996.\ \apj 464:364--72

\bibitem[Stone et al. 1996]{setal96} Stone, J.~M., Hawley, J.~F.,
Gammie, C.~F., \& Balbus, S.~A.\ 1996.\ \apj 463:656--73

\bibitem[Stone \& Pringle 2001]{sp01} Stone, J.~M., \& Pringle,
J.~E.\ 2001.\ \mn 322:461--72

\bibitem[Tassoul 1978]{t78} Tassoul, J.~L.\ 1978.\ {\it Theory of
Rotating Stars,} Princeton Univ.\ Press, Princeton

\bibitem[Terquem \& Papaloizou 1996]{tp96} Terquem, C., \&
Papaloizou, J.~C.B.\ 1996.\ \mn 279:767--84

\bibitem[Triton 1988]{t88} Triton, D.~J.\ 1988.\ {\it Physical Fluid
Dynamics, } Clarendon Press, Oxford

\bibitem[Turner, Stone, \& Sano 2002]{tss02} Turner, N.~J., Stone,
J.~M., \& Sano, T.\ 2002.\ \apj 566:148--63

\bibitem[Uchida \& Shibata 1985]{us85} Uchida, Y., \& Shibata, K.\
1985.\ {\it PASJ} 35:515--35

\bibitem[Umebayashi \& Nakano 1988]{un88} Umebayashi, T., \& Nakano,
T.\ 1988. {\it Prog.\ Theo.\ Phys.\ Suppl.\ } 96:151--60

\bibitem[Varni\`ere \& Tagger 2002]{vt02} Varni\`ere, P., \&
Tagger, M.\ 2002.\ \asap 394:329--38

\bibitem[Velikhov 1959]{v59} Velikhov, E.~P.\ 1959.\ {\it J.\
Expl.\ Theoret.\ Phys.\ (U.S.S.R.)} 36:1398--1404

\bibitem[Vishniac \& Diamond 1989]{vd89} Vishniac, E.~T., \& Diamond, P.\
1989.\ \apj 347:435--47

\bibitem[Ward 1986]{w86} Ward, W.~R.\ 1986.\ {\it Icarus} 67:164--80

\bibitem[Ward 1997]{w97} Ward, W.~R.\ 1997.\ {\it Icarus} 126:261--81

\bibitem[Wardle 1999]{w99} Wardle, M.\ 1999.\ \mn 307:849--56

\bibitem[Wardle \& K\"onigl 1993]{wk93} Wardle, M., \& K\"onigl, A.\
1993.\ \apj 410:218--38

\bibitem[Zahn et al. 1974]{zetal74} Zahn, J.-P., Toomre, J., Spiegel,
E.~A.\ 1974.\ {\it J.\ Fluid Mech.} 64:319--45

\end{thebibliography}
\end{document}